\documentclass[fleqn,usenatbib]{aa}  

\usepackage{txfonts}
\usepackage{graphicx}
\usepackage{ifthen}
\usepackage{url}
\usepackage{amsmath}
\usepackage{bm}
\usepackage{lscape}
\usepackage{rotating}
\usepackage[graphicx]{realboxes}
\usepackage{supertabular}
\usepackage{pdfpages}
\usepackage{footnote}
\usepackage{hyperref}
\usepackage{amssymb}
\makesavenoteenv{tabular}
\makesavenoteenv{table}

\usepackage[T1]{fontenc}
\usepackage{tipa}
\usepackage{ulem}

\hypersetup{
  colorlinks   = true,    
  urlcolor     = blue,    
  linkcolor    = blue,    
  citecolor    = blue      
}

\def\kms{{\rm\,km \  s^{-1}}}

\def\kpc{{\rm\,kpc}}

\def\msun{{\rm\,M_\odot}}

\def\eg{{e.g.{} }}

\usepackage{amsmath}

\def\FeH{{\rm[Fe/H]}}

\begin{document}

   \title{The Pristine Inner Galaxy Survey (PIGS) X}

   \subtitle{Probing the early chemical evolution of the Sagittarius dwarf galaxy with carbon abundances}

\author{Federico Sestito\inst{1,2}
\and Anke Ardern-Arentsen\inst{3}
\and Sara Vitali\inst{4,5}
\and Martin Montelius\inst{6}
\and Romain Lucchesi\inst{7}
\and Kim A. Venn\inst{1}
\and Nicolas F. Martin\inst{8,9}
\and Julio F. Navarro\inst{1}
\and Else Starkenburg\inst{10}
}

\institute{Department of Physics and Astronomy, University of Victoria, PO Box 3055, STN CSC, Victoria BC V8W 3P6, Canada
\and 
Centre for Astrophysics Research, Department of Physics, Astronomy and Mathematics, University of Hertfordshire, Hatfield, AL10 9AB, UK
\email{f.sestito@herts.ac.uk}
\and
Institute of Astronomy, University of Cambridge, Madingley Road, Cambridge CB3 0HA, UK
\and
Instituto de Estudios Astrof\'isicos, Universidad Diego Portales, Av. Ej\'ercito Libertador 441, Santiago, Chile
\and
Millenium Nucleus ERIS
\and
Kapteyn Astronomical Institute, University of Groningen, Landleven 12, 9747 AD Groningen, The Netherlands
\and
Dipartimento di Fisica e Astronomia, Universit\`a degli Studi di Firenze, Via G. Sansone 1, I-50019 Sesto Fiorentino, Italy
\and
Universit\'e de Strasbourg, CNRS, Observatoire astronomique de Strasbourg, UMR 7550, F-67000 Strasbourg, France
\and
Max-Planck-Institut fur Astronomie, Königstuhl 17, D-69117 Heidelberg, Germany
\and
Kapteyn Astronomical Institute, University of Groningen, Landleven 12, 9747 AD Groningen, The Netherlands
}

\date{Received XX; accepted YY}

 
\abstract{ 
We aim to constrain the chemo-dynamical properties of the Sagittarius (Sgr) dwarf galaxy using carbon abundances. Especially at low metallicity, these reveal the early chemical evolution of a system, tracing the supernovae (SNe) that contributed and how much of their ejecta made it into the next stellar generation.
Our sample from the \textit{Pristine} Inner Galaxy Survey (PIGS) includes $\sim 350$ metal-poor ([Fe/H]~$<-1.5$) stars in the main body of Sgr with good quality spectroscopic observations. 
Our metal-poor Sgr population has a larger velocity dispersion than metal-rich Sgr from the literature, which could be explained by outside-in star formation, extreme Galactic tidal perturbations and/or the presence of a metal-rich disc/bar $+$ a metal-poor halo. 
The average carbon abundance [C/Fe] in Sgr is similar to that of other classical dwarf galaxies (DGs) and consistently lower than in the Milky Way by $\sim0.2-0.3$~dex at low metallicity.
The interstellar medium in DGs, including Sgr, may have retained yields from more energetic Population~III~and~II supernovae (SNe), thereby reducing the average [C/Fe]. 
Additionally, SNe~Ia, producing more Fe than C, would start to contribute at lower metallicity in  DGs/Sgr than in the Galaxy. The presence of a [C/Fe] gradient for Sgr stars with $\FeH\gtrsim-2.0$ ($\sim 6.8\times 10^{-4}\ \rm{dex \ arcmin^{-1}}$) suggests that SNe~Ia contributed in the system at those metallicities, especially in its inner regions.
There is a low frequency of carbon-enhanced metal-poor (CEMP) stars in our Sgr sample. At higher metallicity/carbon abundance (mostly CEMP-s) this may be due to photometric selection effects, but those are less likely to affect CEMP-no stars. 
We propose that, given the lower average [C/Fe] in DGs, using the same CEMP definition ([C/Fe]~$>+0.7$) as in the Galaxy  under-predicts the number of CEMP stars in DGs, and for Sgr a cut at [C/Fe]$~\sim +0.35$ may be more appropriate, which brings the frequency of CEMP stars in agreement with that in the Galaxy. 
}


\maketitle
%

\section{Introduction}

The Sagittarius (Sgr) dwarf galaxy \citep{Ibata1994}, located approximately 26.5 kpc  away from us towards the inner Galactic regions  \citep{Vasiliev20}, experienced its first in-fall into the Milky Way (MW) about $5$ Gyr ago \citep[\eg][]{RuizLara20}. As it is being tidally stripped by the MW, its core and two stellar streams are now visible in the Sky \citep{Ibata1994,Mateo98,Majewski03,Law10,Belokurov14}, as well as various associated globular clusters \citep[][]{Sbordone07,Mucciarelli17}. Given its proximity, it is an ideal test-bed for galactic chemo-dynamical models. 

The star formation history (SFH) of Sgr is characterised by multiple star formation episodes, investigated with both high-resolution spectroscopy \citep[\eg][]{Bonifacio00,Monaco05,Chou07,McWilliam13,Hansen18Sgr,Hasselquist17,Hasselquist21,Sestito24Sgr} and photometric  techniques \citep[\eg][]{Bellazzini99,Layden00,Siegel07,Vitali22}. 
So far, studies have typically focussed on metal-rich and relatively young stars, given that they are the prevalent population. Further complicating the study of the oldest/metal-poor stars is the strong overlap in the colour-magnitude diagram between the Milky Way bulge population and stars in Sgr \citep{Monaco05,Mucciarelli17}, especially on the blue, metal-poor side of the red giant branch (RGB) of Sgr. However, the most metal-poor stars are key to understanding the early chemical evolution of Sgr.

An efficient way to discover new members in dwarf galaxies is to use the exquisite Gaia \citep{Gaia16,GaiaEDR3,GaiaDR3} astrometry and photometry alone \citep[\eg][]{Chiti21,Filion21,Yang22,Waller23,Sestito23scl,Sestito23Umi,Hayes23,Jensen24} or to couple it with metal-poor dedicated photometric surveys, e.g. the \textit{Pristine} survey \citep{Starkenburg17b, martin23}, as done in the \textit{Pristine} dwarf galaxy survey \citep[\eg][]{Longeard22,Longeard23}.

Along those lines, the \textit{Pristine} Inner Galaxy Survey (PIGS)  targets metal-poor stars towards the inner regions of the MW \citep{Arentsen20a}, as well as the Sagittarius dwarf galaxy \citep{Vitali22}. The latter work investigated the metallicity distribution of $\sim50,000$ Sgr candidate members as a function of their spatial location, and identified  the largest sample of Sgr candidate members with $\FeH\leq-2.0$ ($\sim1200$ stars). From PIGS, \citet{Sestito24Sgr} followed-up with MIKE high-resolution spectroscopy 12 very metal-poor (VMP, $\FeH\leq-2.0$) Sgr members, the largest and most complete detailed chemical abundance analysis of the VMP Sgr component \citep[vs 4 VMPs in][]{Hansen18Sgr}. The authors interpreted the chemical pattern of the most metal-poor stars as the result of a variety of type II supernovae and asymptotic giant branch stars. A wide range of energetic supernovae and hypernovae with intermediate mass ($10-70\msun$) are needed  to account for the chemical abundances of the lighter elements up to the Fe-peak. The chemical trend of the heavier elements is interpreted as a mixture of yields from compact binary mergers and massive (up to $\sim120\msun$) fast-rotating stars (up to $\sim300\kms$). 

Investigating the origin of carbon in a given stellar population is crucial to understand various astrophysical topics, for example the types of supernovae contributing in a given system, nucleosynthesis in massive stars and binary interaction mechanisms \citep[\eg][]{Frebel07,Vincenzo18,Kobayashi20}. At low metallicity, many stars are found to be carbon-enhanced. Populations of these so-called carbon-enhanced metal-poor (CEMP) stars, with [C/Fe]~$>+0.7$, are powerful probes of the underlying stellar population and the star formation history. Some CEMP stars are thought to carry the imprint of the first generations of supernovae, these are called CEMP-no stars and have sub-solar Ba, [Ba/Fe]~$<0.0$ \citep{Beers05,Aoki07}. It has been suggested that classical DGs have a lower CEMP-no fraction than the MW halo and ultra-faint dwarfs (UFDs) \citep[\eg][]{Starkenburg13, Jablonka15, Kirby15, Simon15, Hansen18, Lucchesi24, Skuladottir15, Skuladottir21, Skuladottir24}. 

Other types of CEMP stars are typically the products of mass transfer from binary interaction with a former asymptotic giant branch (AGB) star companion. These are Ba-rich ([Ba/Fe]~$>+1.0$) due to slow-process channels taking place in the AGB companion and are called CEMP-s stars \citep{Beers05}. The latter group is important to understand the properties of binary populations. In particular, their properties are instructive to understand the nucleosynthetic channels, convection and non-convective processes \citep[\eg][]{Stancliffe07}; the interaction mechanisms, such as the physics of Roche-lobe over-flow and wind accretion \citep[\eg][]{Abate13}; and their influence on the measurement of the velocity dispersion in a system and its dynamical mass \citep[\eg][]{Spencer17,ArroyoPolonio23}, such as its dark matter content. 

From medium-resolution spectroscopy, metallicities and carbon abundances have been measured in only 11 VMP stars in Sgr  \citep{Chiti19,Chiti20Sgr}. In this work, we use the data release of the PIGS low/medium-resolution spectroscopic campaign \citep{Arentsen24} to select the largest sample of low-metallicity ($\FeH\leq-1.5$) Sgr members (356 stars) with measured metallicity, [C/Fe], and radial velocity to date. The dataset and a discussion on the photometric selection effects due to the \textit{Pristine} filter is reported in Section~\ref{sec:data}. The dynamical properties of the metal-rich and metal-poor populations in Sgr are outlined in Section~\ref{sec:rvdist}. A comparison of the [C/Fe] abundances in Sgr with respect the other classical dwarf galaxies (DGs) and the MW halo and inner Galaxy is discussed in Section~\ref{sec:comparison}. We discuss the types and frequencies of CEMP stars in Sgr in Section~\ref{sec:cempnos}, including a suggestion that the definition of CEMP might need revision in DGs.  Conclusions are summarised in Section~\ref{sec:conclusions}.

\section{The \textit{Pristine} Inner Galaxy Survey (PIGS)}\label{sec:data}

PIGS  targets the most metal-poor stars in the inner regions of the Milky Way \citep{Arentsen20a}, using a metallicity-sensitive narrow $CaHK$ filter mounted at Canada-France-Hawaii Telescope (CFHT). Among the photometric metal-poor candidates, $\sim 13\,235$ stars have been observed with the Anglo Australian Telescope (AAT) using the AAOmega+2dF spectrograph. We will refer to them as the PIGS/AAT sample, which is publicly available \citep{Arentsen24}. The AAT setup acquired spectra with low-resolution ($R\sim 1800$) in the blue and with medium-resolution ($R\sim 11\,000$) around the calcium triplet. The analysis is described in detail in \citet{Arentsen20b}, but, briefly, the two arms were fit simultaneously with the \texttt{FERRE} code\footnote{\url{http://github.com/callendeprieto/ferre}} \citep{Prieto2006} to obtain stellar parameters (effective temperature and surface gravity), metallicities, and carbon abundances. The radial velocities (RVs) were derived by cross-correlation of the calcium triplet spectra with synthetic templates. 

\begin{figure*}[h!]
\includegraphics[width=\textwidth]{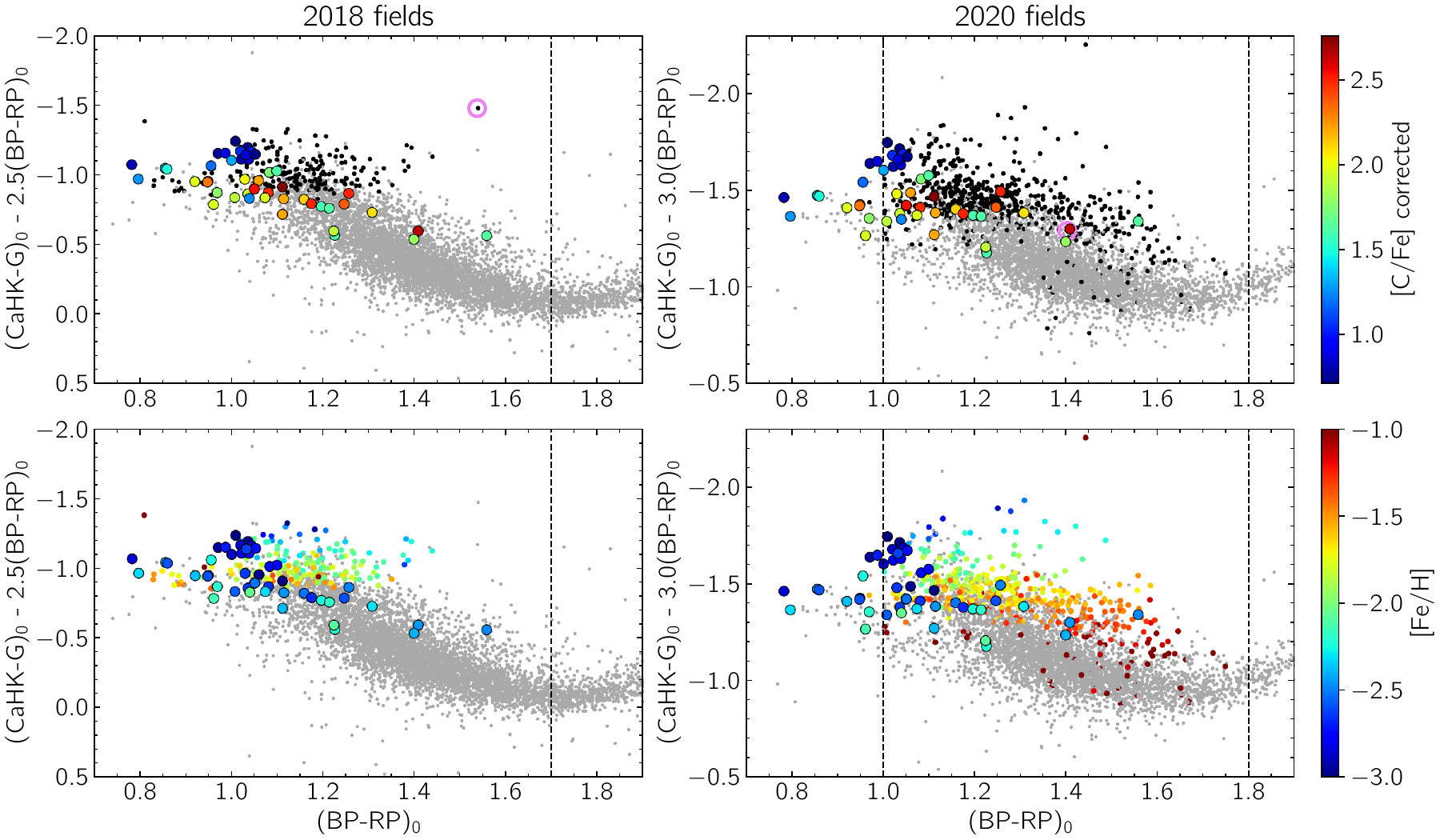}
\caption{Colour-colour diagrams for the two fields observed in 2018 (left) and the two observed in 2020 (right). Grey dots are all stars with PIGS photometry in the targeted fields passing the respective Sgr selection criteria for those years (see Section~\ref{sec:selphot}) and $G<17$. Black dots in the top row are all observed AAT stars in these fields, coloured small dots in the bottom panel are good quality AAT stars coloured by spectroscopic metallicity. Large dots denote \citet{Yoon16} CEMP stars with logg $<2.5$ and $-3<$ [Fe/H] $<-2.0$. Colour coding in the top row is [C/Fe], in the bottom is \FeH{}. Vertical lines indicate colour cuts applied. The stars within the pink circles in the top left and right panels are discussed in Section~\ref{sec:weird} (in the right-hand panel it is at (x,y) $\sim$ (1.4,-1.3)).}
\label{Fig:selection}
\end{figure*}

\subsection{PIGS target selection from photometry}\label{sec:selphot}
Some of the PIGS/AAT fields overlap with the core of the Sagittarius dwarf galaxy and, in four fields, Sgr stars were specifically targeted. Two fields were observed in 2018 and served as a pilot program (\texttt{Field282.0-29.8\_Sag},  \texttt{Field284.0-30.0\_Sag}), two additional fields with more Sgr candidates were observed in 2020 (\texttt{Field282.9-32.1},
\texttt{Field286.0-31.1}). For the 2018 observations, Sgr stars were selected to be within a radius of 0.6  mas yr$^{-1}$  around proper motion $\mu_{\alpha} = -2.7$ mas yr$^{-1}$ and $\mu_{\delta}  = -1.35 $ mas yr$^{-1}$ and parallax $-$ parallax\_error $<$ 0.05 mas. This was relaxed a little in 2020, to a radius of 1  mas yr$^{-1}$ around those proper motions and the parallax $-$ parallax\_error $<$ 0.1 mas. In 2020, suspected variable stars were removed using the flux error and the number of observations \citep{FernandezAlvar21}. Both selections were done using Gaia DR2 \citep{GaiaDR2}. 

The photometric calibration of the PIGS \textit{CaHK} photometry was slightly different when the targets were selected compared to the current, final photometric catalogue, but changes are not expected to be major for the Sgr fields. For the fields from 2018, Sgr candidates  were selected using a horizontal line in $(CaHK - G)_0 - 2.5(BP-RP)_0$ to select the best $\sim100$ Sgr targets per field (and the rest of the AAT fibres were filled with inner Galaxy targets). Observed targets can be seen as black/small coloured points in the \textit{Pristine} colour-colour diagrams in the left-hand panels of Figure~\ref{Fig:selection}, compared to all Sgr candidates in the fields in grey. A red cut at $(BP-RP)_0 = 1.7$ was also made. For the fields in 2020 a different strategy was used, the focus was completely on Sgr and inner Galaxy stars were mostly used as fillers if no fibres could be placed on Sgr stars. Sgr candidates were selected in two ways. The first group contained all stars brighter than $G_0 = 15.5$ and bluer than a [M/H] = $-1.0$ MIST isochrone \citep{Choi16,Dotter16}, this was to get some  red and bright targets that would have been missed in the 2018 selection. The next group contained the most promising metal-poor candidate stars according to \textit{CaHK}, again using a horizontal selection in the colour-colour diagram, this time with factor of 3.0 instead of 2.5 in front of $(BP-RP)_0$. These selections can be seen as black/small coloured points in the right-hand panels of Figure~\ref{Fig:selection}. A colour cut of $1.0 < (BP-RP)_0 < 1.8$ was also made.

\subsection{Selection effects with reference to CEMP stars}\label{sec:seleffect}

Photometric selections of metal-poor stars are plagued by selection effects against carbon-rich stars, especially for cooler stars \citep[\eg][]{Beers99,Rossi05,Goswami06,DaCosta19,Yoon20,Arentsen21,martin23}. This is because carbon has many molecular features in the spectrum, affecting both the narrow-band and broad-band photometry. 

We empirically investigate possible selection effects in our Sgr sample by comparing the location of our observed Sgr/AAT sample in the \textit{Pristine} colour-colour diagram with known CEMP stars from \citet[][hereafter Y16]{Yoon16}. We select giant stars within the relevant Sgr range, making cuts on $\log g < 2.5$ and $-3.0 < \mathrm{[Fe/H]} < -2.0$. Almost all Y16 stars after this cut have $T_\mathrm{eff} > 4500$~K. We use the synthetic $CaHK$ catalogue from \citet{martin23}, derived from Gaia XP spectra \citep{GaiaDR3}, and cross-match it with  Y16 to obtain \textit{Pristine} colour-colour diagram positions for these stars. All $CaHK$ uncertainties for the Y16 stars are less than 0.075~mag, with more than $80\%$ less than 0.05~mag. For the metal-poor regime in the Sgr/PIGS colour-colour diagrams, PIGS $CaHK$ uncertainties are typically less than 0.025~mag.

Large symbols in Figure~\ref{Fig:selection} are CEMP stars from \citet{Yoon16} in the relevant Sgr range. Unfortunately, the Y16 catalogue does not contain many cool giants in this metallicity range, but a small sample of 48 stars remains that can be used. What is clear is that the CEMP stars are mostly not where they are expected to be, given their metallicity -- they are further down in the colour-colour diagrams. A similar conclusion for the \textit{Pristine} survey was reached by \citet{martin23}, who reported that these stars have higher photometric metallicities than their spectroscopic metallicities \citep[see also][]{Caffau20}. Analogously, the SkyMapper survey, which is targeting metal-poor stars with the $v$ filter also in the $CaHK$ region, found a similar bias against CEMP stars, especially for those stars with very large carbon-enhancement  \citep[predominantly CEMP-s,][]{DaCosta19}.

For the 2018 fields (left column of Figure~\ref{Fig:selection}), a large fraction of Y16 CEMP stars falls outside the selected region (y-axis $\lesssim-0.9$). These are mostly stars with $\FeH > -2.5$ and/or [C/Fe] $>+1.8$ -- the regime where CEMP-s stars dominate. Stars with $\FeH < -2.5$ and [C/Fe] $< +1.8$ fall within the selected range -- this combination of \FeH{} and [C/Fe] is in the regime of the Group~II/CEMP-no stars. In the 2020 fields (right column of Figure~\ref{Fig:selection}), more Sgr stars were targeted and the selection boundary lies slightly lower in the colour-colour diagram. More Y16 CEMP stars  now overlap with the selection range, although very much at the edge. The biases are similar to those of the 2018 selection, although a few more stars with $\FeH < -2.5$ and [C/Fe]~$>+2.0$ are included now. From this analysis, we conclude that CEMP-no stars with moderate carbon-enhancement should likely be included in our selection (especially for the 2020 fields, where the majority of our sample comes from), but a large fraction of CEMP-s stars would likely have been excluded. 

Finally, we note that the Y16 sample does not have any stars cooler than 4500~K with $\FeH < -2.5$ or with $\FeH > -2.5$ and [C/Fe]~$<+1.5$. It is therefore difficult to estimate the biases against these stars, although we expect them to be worse for such cool stars. Our analysis in this work is focused on slightly warmer stars so the details of these stars are not crucial. 

\begin{figure*}[h!]
\includegraphics[width=0.95\textwidth]{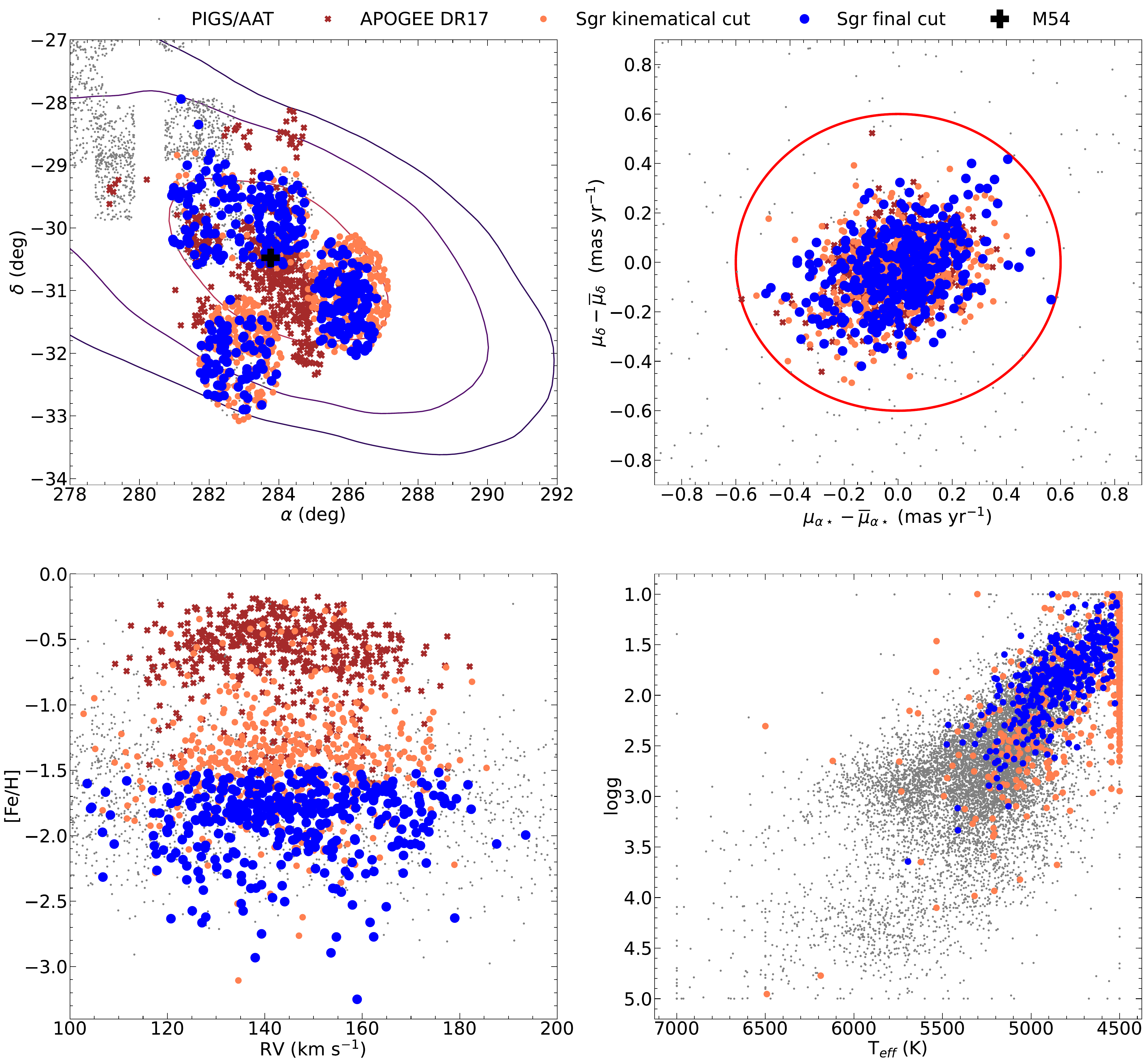}
\caption{Top left panel: On-sky position. Black plus symbol mark the position of the M54 star cluster. Contour lines represent the distribution of Gaia DR3 Sgr candidate members selected on their proper motions as in our kinematical cut. Contour lines  are mark the position at which the number of Sgr candidate members decreases by a factor of 2, 4, and 6, respectively. Top right panel: reduced proper motion space. Bottom left: metallicity$-$radial velocity distribution. Bottom right: Kiel diagram of the PIGS/AAT data, including the Inner Galaxy and Sgr. Blue circles mark the final Sgr cut (356 stars), coral circles denote the Sgr stars from the kinematical selection (834 stars), brown crosses indicate the Sgr members selected from APOGEE DR17, and the  grey dots correspond to the PIGS/AAT sample of the inner Galaxy. Sgr members from APOGEE DR17 have been selected imposing a similar RV and proper motion cut as our sample, a high signal-to-noise ratio in the spectra ($>70$) to ensure good quality in the RV and \FeH{}. APOGEE DR17 stars are not displayed in the Kiel diagram to better highlight  PIGS data.
} 
\label{Fig:onsky}
\end{figure*}

\subsection{Sagittarius spectroscopic sample used in this work}\label{sec:sgrselection}

For this work, to remove the MW contamination from the Sgr candidates, a selection of the Sgr members is made based on the Gaia DR3 proper motions, position on the sky, and radial velocity. In particular, we use the reduced proper motions for Sgr\footnote{$\mu_{\alpha} - \overline{\mu}_{\alpha} = \mu_{\alpha} + 2.69 - 0.009\Delta\alpha +0.002\Delta\delta + 0.00002\Delta\alpha^{3},$ \\ $\mu_{\delta} - \overline{\mu}_{\delta} = \mu_{\delta} + 1.35 + 0.024\Delta\alpha +0.019\Delta\delta + 0.00002\Delta\alpha^{3},$ \\ where $\Delta\alpha,\Delta\delta$ are differences in RA and Dec of each star from the centre of the system ($\alpha_0=283.764$ deg, $\delta_0=-30.480$ deg)}, as defined in \citet{Vasiliev20}. This  takes into account that the proper motion of the members changes as a function of the coordinates. We assume a star to be a Sgr member if it has a reduced proper motion of less than $0.6$ mas yr$^{-1}$ as in \citet{Vasiliev20} and \citet{Vitali22}.  Additionally, Sgr members have RVs in the range from $100$ to $200$ $\kms$ \citep[\eg][]{Ibata1994,Bellazzini2008,Minelli23}. Finally, we limit our analysis to stars with RA $>280^\circ$. This leads to a sample of 834 kinematically selected PIGS/AAT Sgr stars.

Not all the AAT spectra have enough good quality to obtain reliable measurements of [Fe/H] and [C/Fe]. Therefore, bad measurements are removed from the kinematical selection using the flag \texttt{good\_ferre = True}, as suggested in \citet{Arentsen24}. This flag is based on the S/N of the blue spectra, the \texttt{FERRE} $\chi^2$ and the CaT not being double-lined. This further cut leads to 631 Sgr members with available chemistry. The stars with bad S/N in the AAT sample are partly due to issues with the 2dF fibre placement (see discussion in \citealt{Arentsen20b}), which were particularly severe for the two fields observed in 2020 -- this is why the upper/right parts of these fields in RA/Dec (see top-left panel of Fig.~\ref{Fig:onsky}) do not have many stars in the final Sgr cut. 

The stellar parameter grid used in \texttt{FERRE} is limited to $4500 \leq \rm{T_{eff} (K)} \leq 7000$ and $1 \leq \rm{log g} \leq 5$, implying that for stars at the edge of this grid, a wrong model atmosphere might have been adopted to derive the \FeH{} and [C/Fe]. For the Sgr stars, this is particularly an issue at the cool end (see the bottom right panel of Figure~\ref{Fig:onsky}); we, therefore, remove stars with T$_{\rm{eff}} < 4510$~K to avoid stars close to the cool limit of the \texttt{FERRE} grid. For warm stars, the [C/Fe] abundances may not be reliable, we therefore remove stars with T$_{\rm{eff}} > 5700$~K. Because we are interested in the chemistry, we only keep stars with reasonable uncertainties on \FeH{} and [C/Fe] ($< 0.5$~dex). After these cuts, the PIGS/AAT Sgr sample consists of 437 stars. However, in this work we are mainly interested in stars with $\FeH<-1.5$, which results in a final selection of 356 metal-poor PIGS/AAT Sgr members with good measurements of  \FeH{}, [C/Fe], and RV. A table of the Sgr members updated to Gaia DR3 will be available as online material. 

The PIGS/AAT sample ($13\,235$ stars, grey dots), the stars from the kinematical cut (834, coral circles), the final selection (356 stars, blue circles), and Sgr members from APOGEE DR17 \citep[525 stars, brown crosses,][]{APOGEEDR17} are shown in Figure~\ref{Fig:onsky}. The figure displays the position on the sky zoomed in on the Sgr fields (top left panel), the reduced proper motion space (top right), the \FeH{}$-$RV space (bottom left), and the Kiel diagram (bottom right). PIGS/AAT stars in grey dots that lie within the red circle in  proper motion space (top right) do not have  RV compatible with Sgr, and, similarly, PIGS/AAT stars in grey dots with similar RV as Sgr (bottom left) do not match its proper motion. The Kiel diagram clearly shows an overdensity of stars at the cool edge of the \texttt{FERRE} grid, which has been removed as outlined above. Most PIGS/AAT Sgr stars have $1.0 < \log g < 2.5$ and $4500$~K $<$ T$_{\rm{eff}} < 5300$~K.

Part of this work is focused on very carbon-rich objects (Section~\ref{sec:cempnos}), so it is important to be certain that our spectroscopic quality cuts do not bias against such stars. The main quality cuts of relevance are the S/N and the \texttt{FERRE} $\chi^2$. The S/N is determined from the spectra independently of the \texttt{FERRE} fit, in two regions ($4000-4100$~\AA{} and $5000-5100$~\AA), and is not expected to be strongly affected by the carbon abundance, so cutting on it is unlikely to introduce a bias against CEMP stars. If \texttt{FERRE} cannot find a good fit or there are many bad regions in the spectrum, the $\chi^2$ will be high. We inspect all fits of Sgr candidates with bad S/N or bad $\chi^2$ by eye, and identify two clearly carbon-rich stars that are badly fitted, with a high $\chi^2$. Both of these are very cool, very carbon-enhanced and intermediate/very metal-poor, and they will be discussed in Section~\ref{sec:weird}.

\section{On the RV distribution}\label{sec:rvdist}
The RVs of the PIGS/AAT Sgr sample fall within the overall distribution of stars in Sgr's core, ranging between $100-200\kms$ \citep[\eg][]{Ibata1994,Bellazzini2008,Minelli23}, see also Figure~\ref{Fig:onsky}. Various studies have pointed out that the metal-poor population of Sgr, both in the core and in the stream, is more spatially extended and has a larger velocity dispersion $\sigma_{\rm{RV}}$ and a larger systemic velocity $<\rm{RV}>$ than the more metal-rich population \citep[\eg][]{Gibbons17,Johnson20,Penarrubia21,Vitali22,Limberg23,Minelli23}. With the PIGS/AAT Sgr sample, we update these quantities using a more metal-poor, and likely older, population than previous work.

\begin{figure}[h!]
\includegraphics[width=0.5\textwidth]{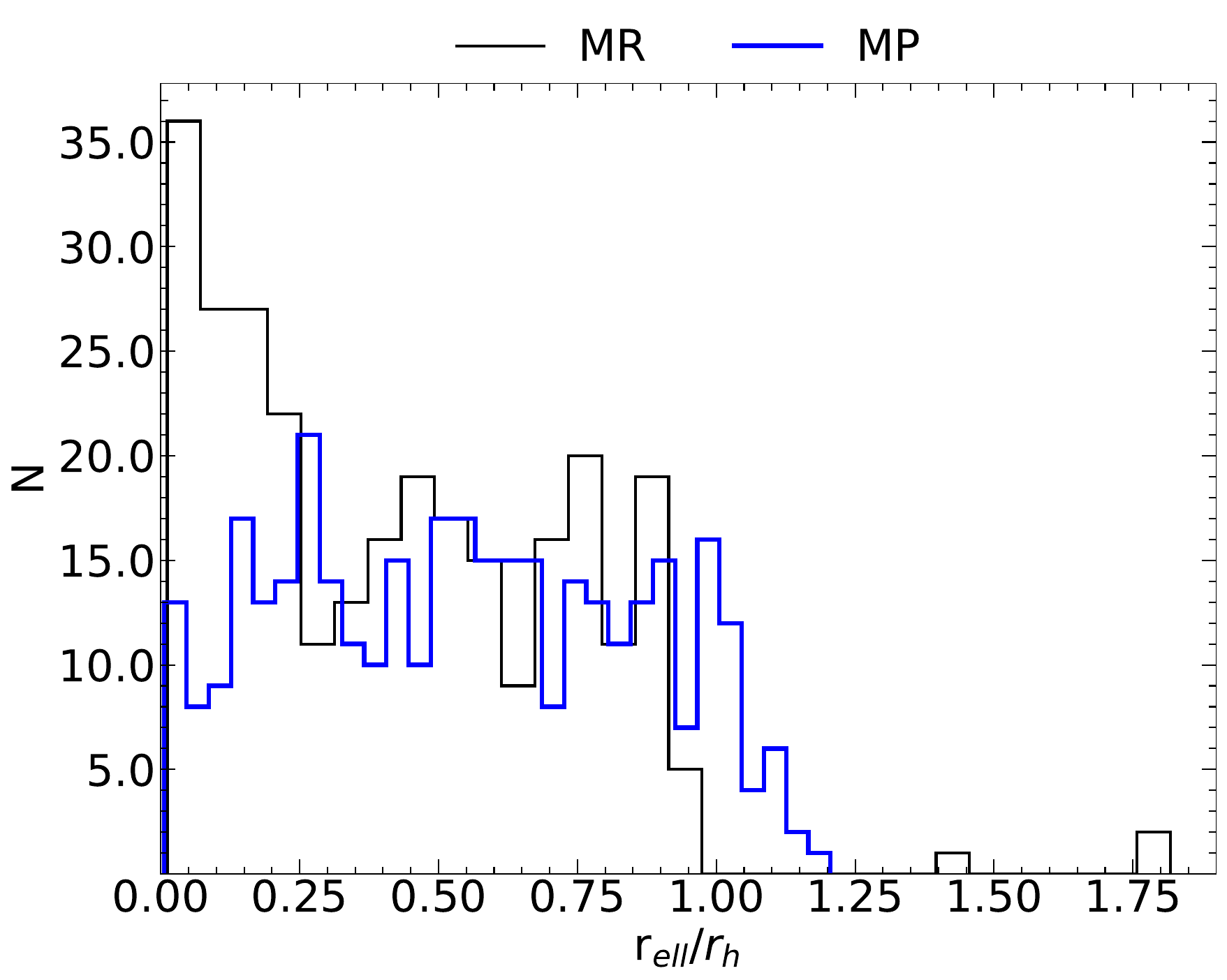}
\caption{Number of stars as a function of the projected elliptical distance. The metal-poor ($\FeH<-1.5$) and metal-rich ($\FeH>-0.6$) are denoted by the blue and black line, respectively. Right ascensions and declinations are converted to tangential plane coordinate assuming the centre of the system as $\alpha_0=283.764$ deg and $\delta_0=-30.480$ deg. The half-light radius is assumed to be $2.6$ kpc \citep{Majewski03,Mucciarelli17} at a distance of $26.5$ kpc \citep{Vasiliev20}, ellipticity $\sim0.57$ and position angle $\sim-104$ deg as in \citet{Vitali22}.}
\label{Fig:hist}
\end{figure}

Sgr stars have been divided into two populations, the metal-poor ($\FeH<-1.5$) from PIGS/AAT and the  metal-rich  ($\FeH>-0.6$) from  APOGEE DR17. The number of stars in these two populations as a function of the projected elliptical distance from Sgr's centre is shown in Figure~\ref{Fig:hist}. The metal-rich population dominates over the metal-poor one in the very inner regions, until a projected elliptical distance of $\sim0.25$ half-light radii ($\rm{r_h}$). Then the two groups from the two surveys are similarly populated.

\begin{figure}[h!]
\includegraphics[width=0.49\textwidth]{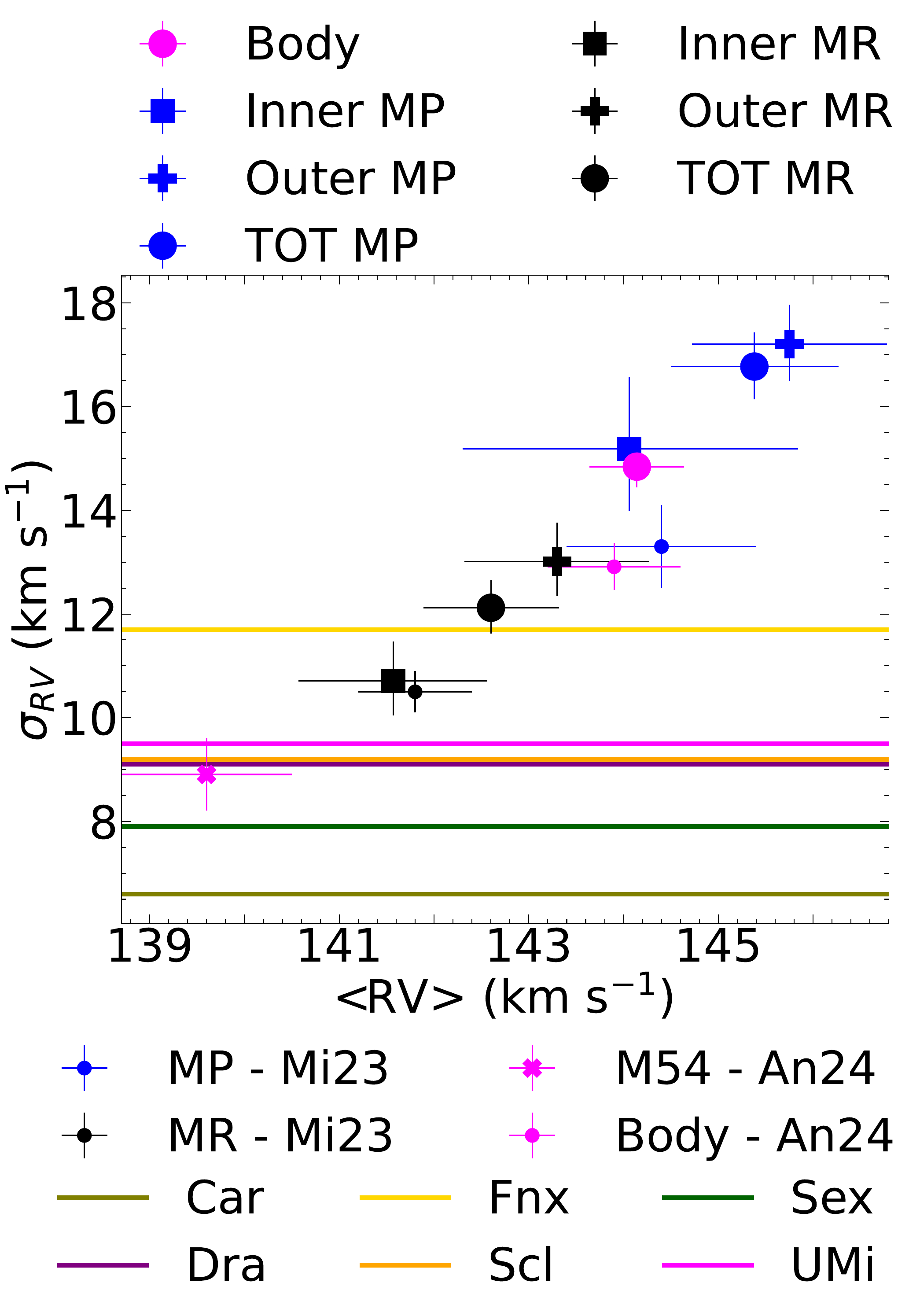}
\caption{RV dispersion vs systemic RV. Large blue and black markers denote the metal-poor ($\FeH\leq-1.5$, MP) stars from Sgr/AAT and the  metal-rich Sgr population ($\FeH>-0.6$, MR) from APOGEE DR17, respectively. The large magenta circle marks the position of the main Sgr's body in this space, considering data from APOGEE and Sgr/AAT. Squares, plusses, and circles correspond to the inner (projected distance $<0.25 \rm{r_h}$), the outer ($\geq0.25 \rm{r_h}$), and the whole population, respectively. A systematic error of 2 $\kms$ is added in quadrature to the RV uncertainties of Sgr/AAT data. The  blue and black small circles mark the MP ($\FeH\leq-0.6$) and MR ($\FeH>-0.6$) populations of Sgr from \citet{Minelli23}, respectively. Magenta cross and small circle denote  M54 and the main body of Sgr as measured by \citet{An24}.  Horizontal solid lines indicate the RV dispersion of the other classical DGs \citep{McVenn2020a}.}
\label{Fig:disp}
\end{figure}

The metal-poor and the metal-rich populations are then divided into two sub-groups  according to their projected elliptical distances: the inner group at  $<0.25\ \rm{r_h}$ vs the outer at $\geq0.25\ \rm{r_h}$. To derive the systemic RV and the RV dispersion, a Bayesian framework embedded in a Monte Carlo Markov chain, based on the Metropolis-Hastings algorithm, is employed. The prior probability distribution  is a step function and it expects these quantities to be in the ranges $90\leq\rm{RV}\leq220\kms$ and $\sigma_{\rm{RV}}\leq40 \kms$. The likelihood is a Gaussian distribution centred on the systemic RV and with a dispersion that takes into account the intrinsic RV dispersion of the system and  the uncertainties of the RV measurements.  The systemic RV, <RV>, vs velocity dispersion, $\sigma_{\rm{RV}}$, are displayed in Figure~\ref{Fig:disp} and reported in Table~\ref{tab:rv}. As reference,  Figure~\ref{Fig:disp} also displays the values for the populations from \citet{Minelli23}, for metal-rich stars ($\FeH>-0.6$, blue small circle) and metal-poor stars ($\FeH\leq-0.6$, but almost no stars with $\FeH < -1.0$, black small circle). 
 We checked for possible systematics between the APOGEE and PIGS radial velocities by comparing both surveys to \textit{Gaia} radial velocities (not limited to Sgr to have many more stars). The difference $\Delta$RV(PIGS $-$ \textit{Gaia}) $= +0.5 \kms$ \citep{Arentsen24} and $\Delta$RV(APOGEE $-$ \textit{Gaia}) $= +0.2 \kms$, implying there is only a $\sim 0.3 \kms$ systematic difference between APOGEE and PIGS.

The overall metal-poor (large blue circle) and metal-rich (large black circle) populations have a systemic RV of $145.4\pm0.9\kms$ and of $\sim142.6\pm0.7\kms$, respectively. These values are compatible with the ones inferred by \citet[][small circles]{Minelli23} adopting a different cut in \FeH{} and a different dataset. The difference in the systemic RV between these populations is significant, given the uncertainties and the precision of the RVs. We did not take into account projection effects.

Recently, \citet{An24} modelled the RV distributions of Sgr and M54, and inferred a difference of $4~\kms$ between M54 (magenta cross marker) and the main body of Sgr (magenta small circle), with mean radial velocities of $139.6 \pm 0.9 \kms$ for M54 and $143.7 \pm 0.7 \kms$ for the main body, with a velocity gradient in the main body. While our sample does not include stars from M54, our estimate of the systemic velocity for the main body is $144.1\pm0.5\kms$ (magenta large circle), which is compatible with the value from  \citet{An24}.
Our results suggest that there is additionally a difference between MP and MR Sgr field populations -- there appears to be an increasing mean RV going from M54, to metal-rich field stars, to metal-poor field stars.

In agreement with previous work on  the stream and core \citep[\eg][]{Gibbons17, Johnson20, Penarrubia21, Vitali22, Limberg23, Minelli23}, we find that the overall metal-poor population has a velocity dispersion larger than the metal-rich counterpart, in our case $\sigma_{\rm{RV}}\sim 17\kms$ vs $\sigma_{\rm{RV}} \sim 12 \kms$, respectively. Also to be noted from Figure~\ref{Fig:disp}: the inner populations in our analysis (large squares), both MP and MR, have lower RV dispersion and lower systemic RV than their respective outer populations (large plus markers). For all the populations and subgroups, the velocity dispersion and the systemic velocity are found  to be considerably higher than the values for M54 (magenta cross marker). The latter has been classified as a nuclear star cluster, which might explain its higher velocity dispersion compared to isolated globular clusters.  Its velocity dispersion might have been inflated by the dark matter halo of Sgr, by tidal interactions, and by the multiple burst of star formation \citep[\eg][]{Carlberg22,Kacharov22,Herlan23,Gray24}. The MP and MR populations should not be contaminated by many M54 members.

\begin{table}
\caption[]{Systemic RVs and RVs dispersions.}
\centering
\resizebox{0.5\textwidth}{!}{
\begin{tabular}{lccc}
\hline
Population  & <RV> & $\sigma_{\rm{RV}}$ & Dataset \\
 & ($\kms$)  & ($\kms$) &   \\ \hline \\[0.05cm]

MR$-$Inner & $ 141.6 ^{+ 1.0 }_{- 1.0 }$ & $ 10.7 ^{+ 0.8 }_{- 0.7 }$ & APOGEE \\[0.2cm]
MR$-$Outer & $ 143.3 ^{+ 1.0 }_{- 1.0 }$& $ 13.0 ^{+ 0.8 }_{- 0.7 }$& APOGEE \\[0.2cm]
MR$-$Tot & $ 142.6 ^{+ 0.7 }_{- 0.7 }$ & $  12.1 ^{+ 0.5 }_{- 0.5 }$ & APOGEE \\[0.2cm]

MP$-$Inner & $144.1 ^{+ 1.8 }_{- 1.8 }$  &  $15.2 ^{+ 1.4 }_{- 1.2 }$ & AAT \\[0.2cm]
MP$-$Outer & $ 145.8 ^{+ 1.0 }_{- 1.0 }$ & $ 17.2 ^{+ 0.8 }_{- 0.7 }$ & AAT \\[0.2cm]
MP$-$Tot & $ 145.4 ^{+ 0.9 }_{- 0.9 }$ & $ 16.8 ^{+ 0.7 }_{- 0.6 }$& AAT \\[0.2cm]

Body & $ 144.1 ^{+ 0.5 }_{- 0.5 }$ & $ 14.8 ^{+ 0.4 }_{- 0.4 }$ & AAT +\\  & & & APOGEE \\[0.2cm]
\hline
\end{tabular}
}
\tablefoot{Systemic RVs and RVs dispersions for the metal-poor and metal-rich populations and for the whole body. The values for the inner, outer, and whole groups are reported, together with the source of the datasets.}
\label{tab:rv}
\end{table}

\subsection{Internal and external mechanisms in play}

Various internal and external mechanisms can  affect the chemo-dynamical properties of a system. For instance, the internal morphology can play a role.  In this regard, a dynamically hotter MP  and a colder MR population with weak rotation has been proposed to indicate the presence of a metal-rich thick and rotating disc or bar surrounded by a more dispersed and metal-poor stellar halo  in Sgr \citep[][]{Mayer01,Sanchez10,Kazantzidis11,delPino21,Carlberg22,Minelli23,Lokas24}. {Both observations and simulations suggest that the rotating bar should have a length of $2-2.5\kpc$ \citep{delPino21,Lokas24}, which  correspond  to an elliptical radius of $0.8-1.0 \rm{r_h}$. As shown in Figure~\ref{Fig:hist}, the majority of the stars from both APOGEE and PIGS lies within 1 half-light radius. The presence of such a rotating disc/bar would also explain some chemo-dynamical properties of the stellar streams associated with Sgr \citep{Penarrubia10,Oria22,Carlberg22}. The fact that the MR population, either in the inner or in the outer regions, has a lower velocity dispersion and a lower systemic RV than the MP supports the idea that these two groups populate two different structures, such as a ``disc/bar'' and a stellar ``halo'' of Sgr. If so, projection effects on the bar are another ingredient  explaining the different systemic RV from the MP group.

Additionally, outside-in star formation has been proposed as one mechanism to explain the different spatial and kinematical properties between MP and MR populations in DGs, such as the gradient in the velocity dispersion  \citep[\eg][]{Tolstoy04, Battaglia06, Battaglia08b, Zhang12, Hidalgo13, BenitezLlambay16, Revaz18, Sestito23scl, Sestito23Umi, Tolstoy23}. In this scenario, the oldest MP population would form spatially everywhere in the system, and their supernovae would enrich the ISM. Then some of the gas might have sunk to the inner region with time, forming younger and more metal-rich stars that are more gravitationally bound to the system. As a result of this, the MP population would be more spatially extended and kinematically hotter than the MR one, with the latter being confined mostly to the inner regions with a lower velocity dispersion. 

The main external mechanisms that can affect the dynamical properties of a DG are merging events and tidal stripping. 
In case of the former, stars will be heated up by the accreted system, and likely the less bound ones, such as in the outskirts, will be more affected. Then, the additional gas from the accreted system (if it has any) can sink into the inner regions, triggering the formation of new stars that are more metal-rich \citep{BenitezLlambay16}. 
In addition, tidal stripping also influences the distribution and kinematics of the outskirts, which are less bound, of a system. In fact, the ongoing stripping of Sgr resulted in the formation of the Sgr stellar streams, which are known to be more metal-poor on average than the core \citep[\eg][]{Hayes20,Limberg23,Cunningham24}. It has been proposed that Sgr has interacted gravitationally with the MW for more than $8$ Gyr, with its first pericentric passage likely to have happened around $5-6$ Gyr ago \citep{Law10,RuizLara20}. The MR population in Sgr has an estimated age spanning from 4 to 8 Gyr -- their star formation, or part of it, might have have been triggered by Galactic perturbations at, or close to, the first pericentric passage. Investigations on simulated galaxies reveals that the extreme tidal effects that Sgr is undergoing might have affected the system's morphology, e.g. it could have reshaped its disc (if it had one) into a prolate rotating bar structure \citep{Lokas24}.

\subsection{Comparison to other DGs}

The values of the  velocity dispersion for the other 6 classical DGs \citep[horizontal lines,][]{McVenn2020a} are also reported in Figure~\ref{Fig:disp} as a reference. The velocity dispersion for the MR population in Sgr is similar to Fornax's value, which is the highest among the DGs compilation. The $\sigma_{\rm{RV}}$ for the MP population in Sgr is significantly higher than the averages for the other DGs. This could be due to an observational bias, such as the $\sigma_{\rm{RV}}$ in the reference galaxies are calculated from the overall population, which is mostly more metal-rich than the MP population in Sgr. As an example, the velocity dispersion for the overall population in Sculptor is around $7\kms$, while restricting to the more dispersed metal-poor stars would provide a $\sigma_{\rm{RV}}\sim10-12\kms$ \citep{Tolstoy04,Battaglia08b,Walker11,Tolstoy23,Sestito23scl}. In addition, Sgr has a total mass higher than the other DGs reported in Figure~\ref{Fig:disp} and it is experiencing strong Galactic tidal stripping, which is far more extreme than in the other systems \citep[\eg][and references therein]{Battaglia22,Pace22}, which both concur to inflate the $\sigma_{\rm{RV}}$ of this system.

\section{Carbon trends in Sagittarius}\label{sec:comparison}

We next focus our attention on the chemistry of Sgr, specifically the abundance of carbon. As discussed in the Introduction, carbon abundances can trace the early chemical evolution of a system \citep[\eg][]{Frebel07,Vincenzo18,Kobayashi20}. Is the level of carbon in Sgr similar to that in the other classical DGs? What about in comparison with the inner Galaxy and the MW halo? To answer these questions, in Figure~\ref{Fig:cfe_comp} we present the average [C/Fe] ratio as a function of the metallicity for Sgr (red circles) compared to the classical DGs (left panel) and compared to the inner Galaxy and the MW halo (right panel). All carbon abundances have been corrected for evolutionary effects according to \citet{Placco14}, see \citet{Arentsen21} for details. We find that the Sgr carbon abundance slightly rises with decreasing metallicity. 

\begin{figure*}[h!]
\includegraphics[width=\textwidth]{plots/cfe_comp4.pdf}
\caption{Average [C/Fe] vs \FeH{} divided into 7 metallicity bins ($\sim0.25$ dex). CEMP stars have been removed and carbon abundances have been corrected for evolutionary effects according to \citet{Placco14}. Left panel: comparison with classical DGs (coloured markers), Car, Dra, Fnx, Scl, Sex, and Umi are  from \citet{Lucchesi24}, while LMC data is from \citet{Chiti24} and \citet{Oh24}.  [C/Fe] in classical DGs for which there are less than 2 stars are not displayed. Right panel: Comparison with the MW. MW halo stars (black markers) are from  \citet{Aguado19}, revised as in \citet{Arentsen22}. Inner Galaxy from PIGS/AAT \citep[grey markers,][]{Arentsen24} are divided into three groups, the whole  sample (grey circles, solid line), the stars confined into the inner regions (grey crosses, dash-dot line, apocenter $<3$ kpc) and the halo interlopers (grey plusses, dash-line, apocenter $>8$ kpc). Bins populated by less than 5 stars are removed. MW stars from PIGS are selected to have $\log g<2.3$, while compilation from \citet{Aguado19} is restricted to stars with $\log g<3.0$. In both cases,  AGBs are removed. [C/Fe] ratios from all the datasets are corrected for the evolutionary effects as in \citet{Placco14}. An offset of up to $\pm0.05$ is added to the metallicity bins of the MW and DGs compilations to better display the markers and the uncertainties on the average [C/Fe].}
\label{Fig:cfe_comp}
\end{figure*}

\subsection{Halo and inner Milky Way}

There are a number of studies that have explored the carbon abundance of low-metallicity stars in the MW halo, and to a lesser extent in the inner Galaxy. \citet[][]{Arentsen22} showed that  trends involving carbon abundances are very sensitive to the assumptions made in the synthetic spectroscopic grids (e.g. the model atmospheres, the adopted atomic and molecular data) and the employed pipeline, with large systematic offsets between different literature samples (see their Figure~4). To not bias our conclusions, in the comparison with the halo and the inner Galaxy, we restrict ourselves to [C/Fe] measured within PIGS and the \textit{Pristine} survey, which have all been derived with the same methodology.

The inner Galaxy PIGS/AAT sample is selected from \citet{Arentsen24}, restricted to those stars with good measurements of stellar parameters, metallicities and carbon abundances, as in our Sgr sample. An additional cut is imposed to select stars with similar surface gravity as the bulk of the Sgr sample ($\log g<2.3$) and to remove the region of early asymptotic giant branch stars (eAGBs) whose carbon abundances have been altered by stellar evolution \citep{Arentsen21}. This selection is composed of 2318 stars with $\FeH<-1.5$ (grey circles in the right-hand panel of Figure~\ref{Fig:cfe_comp}). Additionally, this sample is split into two sub-groups according to their Galactic apocentric distances, those that remain confined in the inner Galaxy (apocentre $<3$ kpc, grey crosses) and the ``halo interlopers'' (apocentre $>8$ kpc, grey plus markers). The former and the latter are composed of 1032 and 276 stars, respectively. For the MW halo, we include the  \textit{Pristine} medium-resolution follow-up sample from \citet[][141 stars, black plus markers]{Aguado19}, with carbon abundances corrected for spurious $\log g$ determinations following \citet{Arentsen22}. This sample has a less  restrictive cut on the surface gravity, namely $\log g <3.0$.

Although the same trend is visible for the Milky Way and Sgr samples, namely a rise in carbon abundance with decreasing metallicity, the average carbon abundances are higher in the Milky Way samples compared to Sgr, and the rise appears to be less steep in Sgr. The [C/Fe] difference between Sgr and the Milky Way starts at $\sim 0.1$~dex for $\FeH = -1.6$ and increases to $0.3-0.4$~dex for $\FeH < -2.5$. 

The average carbon abundance of the MW (inner regions and the halo) is also higher than most of the classical DGs, except for Fornax (Fnx, gold squares). The difference in carbon abundances can be interpreted as a different population of SNe~II and AGB stars that contributed to the chemical enrichment of the dwarf galaxies in comparison with the one of the Galaxy. In  particular, a higher contribution of faint and core-collapse SNe could provide a higher [C/Fe] ratio \citep{Umeda03,Limongi03,Iwamoto05,Kobayashi06,Kobayashi20,Vanni23}.

The physical and chemical properties of the building blocks that contributed to the formation of the proto-Galaxy are still under discussion \citep[\eg][]{Schiavon17,Helmi20,Santistevan21,Sestito21}, as well as the importance of an ancient in-situ component \citep{Belokurov22,Belokurov23}. Did the early building blocks have a chemical evolution similar of the present UFDs? What about their masses and sizes, or, in other words,  are the building blocks comparable to classical DGs or to smaller UFDs \citep[see][]{Deason16}?

For the PIGS inner Galaxy sample, there is a slight difference in the average level of carbon abundance between the ``confined'' (plusses, lower [C/Fe]) and ``halo interloper'' (crosses, higher [C/Fe]) samples, of the order of 0.05-0.10~dex. This could potentially be connected to different building blocks contributing to these populations, e.g. more chemically evolved ones to the confined population and more chemically pristine systems to the halo population (see also the discussion in \citealt{Arentsen24}). We further discuss the connection to dwarf galaxies and their chemical evolution in Section~\ref{sec:dgs}.

\subsection{Note on possible systematics}

As previously discussed, the PIGS/AAT inner Galaxy and the Sgr stars have been analysed with the same methodology applied to the same AAT spectra, and the \citet{Aguado19} sample has been analysed with the same methodology as well, so systematic differences should hopefully be minimal. One caveat here is that [$\alpha$/Fe] is fixed in the analysis, to $+0.4$. However, various high-resolution spectroscopic works showed that the majority of the inner Galaxy VMP stars have similar [$\alpha$/Fe] compared to typical halo stars \citep{Howes14,Howes15,Howes16,Sestito23}, and the $\alpha-$abundances are also very similar between the MW and Sgr in the VMP regime \citep{Hansen18Sgr,Sestito24Sgr}. Therefore, we should not expect significant biases in the \texttt{FERRE} analyses due to [$\alpha$/Fe] differences.

We note that the magnitude of the evolutionary carbon correction following \citet{Placco14} also depends on the natal nitrogen abundances of stars, which may  differ for each formation site, but are all assumed to be [N/Fe]~$= 0.0$ in the calculations. However, the predicted effect on the carbon corrections is much smaller than the difference we find between Sgr and the Milky Way -- Figure~1 of \citet{Placco14} shows that for a $\FeH = -2.3$ star, the difference in the carbon correction between a [N/Fe] of $-0.5$ and $+0.5$ at birth is at most $\sim0.05$~dex. Therefore, a different average level of [N/Fe] between the MW and Sgr would not impact our findings. The evolutionary corrections may also potentially be better or worse in some parts of the parameter space (e.g. depending on $\log g$), so it is crucial to compare stars in similar evolutionary phases. We attempted this by limiting the reference samples in $\log g$, but the distributions of evolutionary phases are not exactly the same. 

What might be the effect of photometric selection effects on trends of carbon? As discussed previously, very carbon-rich stars are likely excluded from our selection because they look too metal-rich. Could our selection be biased even for ``carbon-normal'' stars, selecting only those with relatively lower carbon abundances? This is unlikely to be the case, especially for $\FeH < -2$, given that the carbon features are relatively weak for carbon-normal VMP stars and given that our selection was not only targeting VMP stars, but also probed the slightly more metal-rich population. 

Finally, we checked potential systematics on the mean [C/Fe] and its trend with metallicity as a function of the surface gravity. As a sanity check, we repeated the exercise of Figure~\ref{Fig:cfe_comp}, restricting the Sgr and MW compilations to stars with $1.8<\log g <2.3$ (for lower $\log g$, the \citealt{Placco14} evolutionary carbon corrections become more important). We find no qualitative or quantitative differences between this more strict cut and the one applied to produce Figure~\ref{Fig:cfe_comp}. However, we note that the MW halo sample from \citet{Aguado19} would not have enough stars to populate all the metallicity bins for this limited $\log g$ selection. 

\subsection{Dwarf galaxies}\label{sec:dgs}

To compare the average [C/Fe] of Sgr with classical DGs, stars with $\FeH\leq -1.5$  have been selected from the DG members summarised in \citet{Lucchesi24}, \citet{Chiti24}, and \citet{Oh24}. The compilation from  \citet{Lucchesi24} is composed of 442 stars (16 CEMP-no) and distributed in 7 classical DGs, namely Canes Venatici I (CVn~I, 1 star, \citealt{Yoon20}), Carina (Car, 8 stars, \citealt{Venn12,Susmitha17,Lucchesi24}), Draco (Dra, 161 stars, \citealt{Kirby15a}), Fornax (Fnx, 14 stars, \citealt{Tafelmeyer10,Kirby15a,Lucchesi24}), Sculptor (Scl, 173 stars, \citealt{Kirby12,Kirby15a,Skuladottir15,Skuladottir24}), Sextans (Sex, 4 stars, \citealt{Tafelmeyer10,Lucchesi20}), Ursa Minor (UMi, 81 stars, \citealt{Kirby12,Kirby15a}). The compilations from \citet{Chiti24} and \citet{Oh24} include members of the Large Magellanic Cloud (LMC) for a total of 21 stars (no CEMP). The systems from these compilations, excluding CVn~I and CEMP-no stars, are displayed in Figure~\ref{Fig:cfe_comp} with coloured circles, diamonds,  squares, and plusses. 

The average level of [C/Fe] in Sgr is within the wide range of the 7 classical DGs. In particular, the average carbon abundance in Sgr appears to be higher than  in Scl for $\FeH>-2.4$ by up to $\sim0.3$ dex. Compared to Car, Sgr's [C/Fe] level is also higher, for $\FeH\lesssim-2.4$ by at least $\sim0.3$ dex. As proposed by \citet{Skuladottir24}, the strikingly low amount of [C/Fe] in Scl and Car might be explained by a strong imprint of hypernovae from Pop~III stars.  Thus, classical DGs and stars with such low carbon level might be crucial for understanding the energy distribution of the primordial generation of stars \citep[\eg][]{Koutsouridou23}. 

Another nucleosynthetic channel that contributes to lower the [C/Fe] is from SNe~Ia, in which the production of Fe exceeds that of C \citep{Iwamoto99}. This event might be responsible for lowering the [C/Fe] in Dra and UMi for $\FeH\gtrsim-2.5$, as also shown in \citet{Kirby15}. Chemical abundance analysis from \citet{Cohen09} reveals that the level of [C/Fe] in Dra strongly decreases around $\FeH\sim-2.5$, such as the  metallicity at which SNe~Ia starts to kick in. Similarly, \citet{Sestito23Umi} discovered that the contribution of SNe~Ia in UMi starts at $\FeH\sim-2.1$. In Sgr, the contribution of SNe~Ia is absent in the  VMP regime. However, \citet{Sestito24Sgr} suggest that the trend of [Co/Fe] at $\FeH\gtrsim-2.0$  might be an indication of a possible contribution of SNe~Ia in Sgr. This can also explain the lower [C/Fe] at $\FeH\gtrsim-2.0$ compared to the more metal-poor bins. A more thorough investigation of this metallicity regime in Sgr will be explored by PIGS in a coming paper (Vitali et al., in prep.).

\citet{Sestito24Sgr} discussed the early chemical enrichment phase of Sgr from the detailed chemical abundances of 11 VMP stars. The chemical pattern of Sgr stars has been interpreted as the result of a mixture of Pop~III~and~II stars contributing \citep{Sestito24Sgr}. In particular,  intermediate-mass high-energy and hypernovae are needed to explain the abundance patterns of the lighter elements up to the Fe-peak, while compact binary merger events and fast-rotating (up to $\sim300\kms$) intermediate-mass to massive metal-poor stars ($\sim25-120\msun$) are needed to account for the level of the heavy elements. No evidence for contributions from pair-instability supernovae has been found in \citet{Sestito24Sgr}. This mixture of various energetic SNe events appears to be common in classical DGs, and therefore explain the similarity in [C/Fe] between these systems and their lower level compared to the MW, see Section below for a further discussion on this topic.

\subsection{The different supernovae enrichment}\label{sec:super}

The different amount of [C/Fe] among the classical DGs and their lower level compared to the MW can be interpreted as the imprint of a  different chemical evolution and a different efficiency in retaining the ejecta of SNe. 
For instance, the chemical evolution models from \citet{Vanni23} suggest that DGs would have been polluted by a mixture of SNe~II from Population~III~and~II stars vs a more pristine population of SNe~II in the building blocks of the MW halo \citep[see also][]{Skuladottir24}. The higher fraction of Pop~II would have contributed to partially lower the average [C/Fe] \citep{Vanni23}.

In addition, the ISM of classical DGs is considered to be homogeneously mixed, therefore able to have retained the ejected yields from the most energetic events \citep{Skuladottir24}, such as high-energy SNe~II, hypernovae, and potential pair-instability SNe~II. The retention of the ejected yields from the most energetic events would lower the average amount of [C/Fe], given they would  produce more Fe than C \citep[\eg][]{Limongi18,Kobayashi20,Koutsouridou23,Vanni23}. 

While there is a consensus that massive systems would contribute to the formation of the MW \citep[\eg][]{Deason16}, it is still an open question whether the MW's building blocks resembled UFDs or DGs in terms of their ISM efficiency in retaining SNe yields or regarding their star formation history or their initial mass function. We interpret the higher average [C/Fe] of the MW  as an indication that the ISM efficiency of the MW's building blocks is similar to UFDs, hence unable to retain the most energetic events \citep[\eg][]{Ji16c,Roederer16b,Hansen17,Kobayashi20,Applebaum21,Waller23,Sestito24gh}. 
Therefore, the ISM of the building blocks of the MW, should be the fossil of the lower energetic events only \citep{Koutsouridou23,Vanni23,Skuladottir24}. Additionally, if inhomogeneous chemical enrichment is in place, asymptotic giant branch stars (AGBs)  can also be an extra source for the level of carbon, even at lower metallicities \citep{Kobayashi14,Vincenzo18,Kobayashi20}. 

Figure~\ref{Fig:cfe_comp} also shows a  difference in the average [C/Fe] between the MW halo and the inner Galaxy, especially those stars confined within 3~kpc. Recently, \citet{Pagnini23} suggested that a potential dearth of CEMP stars in the inner Galaxy could be due to the very high star formation rates at early times. The star formation would be so intense that stars massive enough to explode as pair-instability SNe would form, which would lower the average [C/Fe] compared to the halo. However, no star carrying the imprint of pair-instability SNe has been found so far in the Galaxy \citep[\eg][]{Lucey22,Sestito23,Skuladottir24b}.

Furthermore, SNe~Ia can  concur to lower the average [C/Fe] in a given system \citep{Iwamoto99}. The contribution of SNe~Ia might start at $\FeH\gtrsim-2.5$ in some classical DGs \citep[\eg][]{Cohen09,Venn12,Kirby15,Sestito23Umi}, and likely between $-2.0\lesssim\FeH\lesssim-1.5$ for Sgr \citep{Sestito24Sgr}. This is not the case for the MW, where SNe~Ia starts to kick in at higher metallicities, $\FeH\sim-1.0$ \citep[\eg][]{McWilliam97,Matteucci03,Venn04}. Therefore, the lower average [C/Fe] at $\FeH\gtrsim-2.5$ in DGs and at $\FeH\gtrsim-2.0$ in Sgr can also be caused by the contribution of SNe~Ia.

\subsection{The radial gradient of [C/Fe]}\label{sec:carbongrad}

Our sample is large enough and covers enough of Sgr to test whether there may be any radial gradients in [C/Fe]. To avoid potential systematic effects in [C/Fe] between radial bins due to differences in stellar parameter coverage, we limit the sample to $1.8< \log g < 2.3$ for this analysis. We find that the general picture of our results does not change compared to using a more generous cut or the full sample, but the behaviour is cleaner for the limited sample.

The median [C/Fe] as a function of the projected elliptical distance is shown in Figure~\ref{Fig:cfe_ell}. The Sgr PIGS/AAT sample is divided into two sub-groups, the low-metallicity (blue circles, $-2.5\leq\FeH\leq-2.0$) and a slightly more metal-rich group (navy circles, $-2.0 < \FeH \leq -1.5$), and removing CEMP stars from the calculations. 
There is a net positive [C/Fe] gradient for the slightly more metal-rich sub-group, with a difference of $\sim+0.25$ dex between the very inner region and the outskirts of Sgr. This leads to a positive gradient in [C/Fe] of about $\nabla \rm{[C/Fe]} \sim 0.23\ \rm{dex \ r_h^{-1}}$ or
$ \sim 8.8\times 10^{-2}\ \rm{dex \ kpc^{-1}}$ or $\sim 6.8\times 10^{-4}\ \rm{dex \ arcmin^{-1}}$.

Regarding the low-metallicity sub-group, a mild positive gradient is visible if the innermost bin is not considered. In this case, the difference in [C/Fe] would be $\sim0.2$ dex between the inner to the outer Sgr's regions. To be taken into account, uncertainties on the average [C/Fe] are larger for the low-metallicity sub-group than the more metal-rich one.

\begin{figure}
\includegraphics[width=0.5\textwidth]{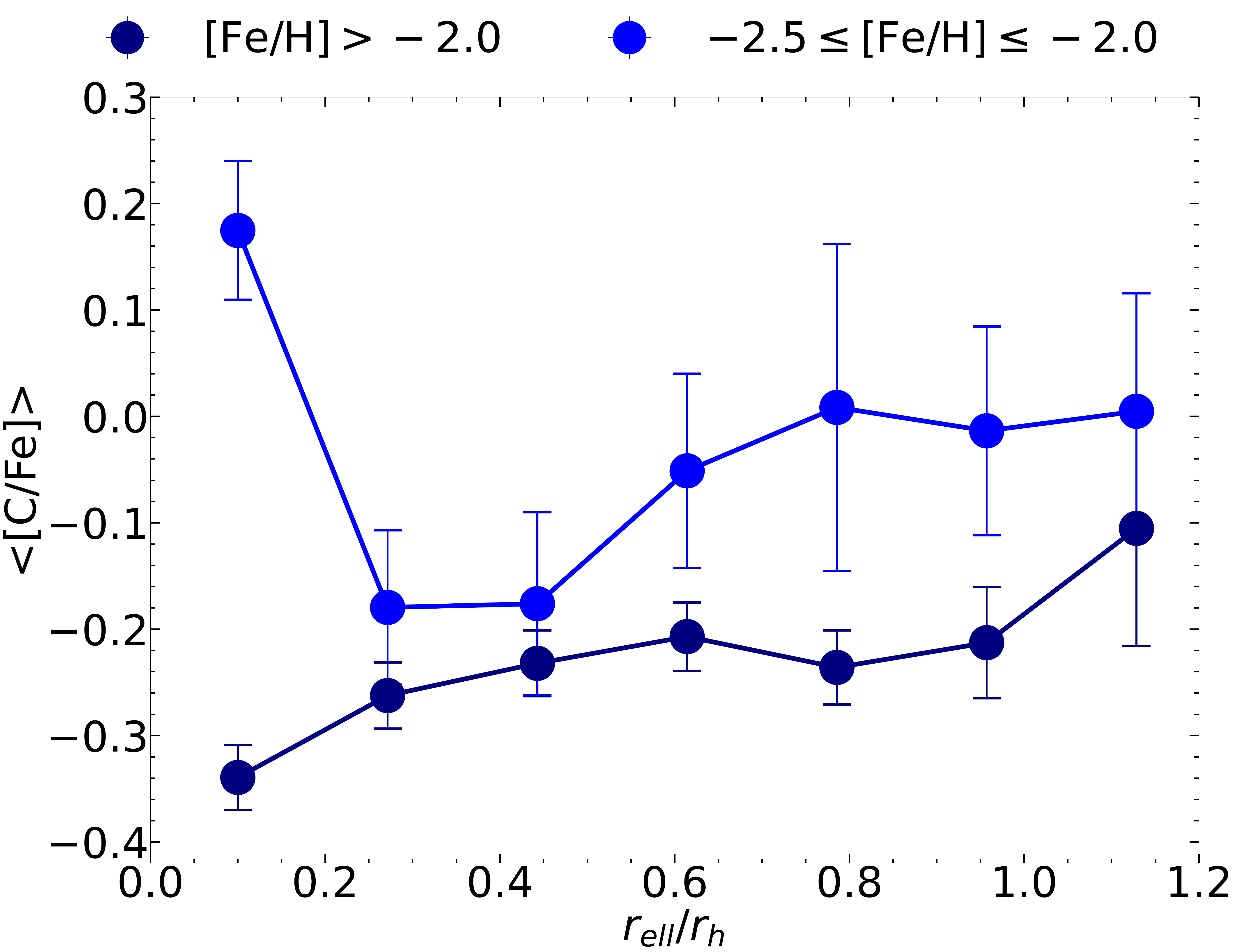}
\caption{Median [C/Fe] as a function of the  projected elliptical distance. Stars from the final selection of Sgr PIGS/AAT. The median is obtained removing the sample from CEMP stars and dividing it into distance bins and into two sub-groups, the more metal-poor (blue circles, $-2.5\leq\FeH\leq-2.0$) and the slightly more metal-rich (navy circles, $\FeH>-2.0$). Stars have been selected to have $1.8<\log g<2.3$.}
\label{Fig:cfe_ell}
\end{figure}

Is the more pronounced gradient at higher metallicities connected to a different chemical enrichment between the two populations? A couple of concurrent mechanisms might explain these gradients: outside-in star formation and the contribution of SNe~Ia. 

The former, as discussed in Section~\ref{sec:rvdist}, implies that the oldest and most metal-poor stars should form everywhere in the system and would carry a similar imprint of nucleosynthetic events, if also homogeneous mixing applies to the system. In the case that the ISM is not completely homogeneously mixed between the inner regions and the outskirts, these two regions might carry different level of [C/Fe]. Likely, the outskirts would be less efficient in retaining the more energetic events as the inner regions, resulting in a higher average [C/Fe].

The stellar feedback from the first supernovae would expel the gas outside the system, which then later would be re-accreted onto the inner regions, where slightly more metal-rich stars would form. 
These relatively more metal-rich inner stars might carry the imprint of SNe~Ia as well. As discussed in Section~\ref{sec:super}, SNe~Ia can lower the average [C/Fe] \citep{Iwamoto99}, and the higher contribution of these events in the inner regions  would explain the positive gradient in [C/Fe]. This result would be an indication, in addition to the trend of [Co/Fe] in \citet{Sestito24Sgr}, that SNe~Ia might have started to kick in in Sgr  at metallicities between $-2.0<\FeH<-1.5$, well below what was previously inferred \citep[$\FeH\sim-1.27$, \eg][]{deBoer14}.

\section{CEMP stars}\label{sec:cempnos}

As discussed in the Introduction, CEMP stars are of interest because they probe the properties of the First Stars and early chemical evolution (CEMP-no) and of binary populations (CEMP-s). Next, we investigate the properties of CEMP stars in Sgr with the PIGS/AAT Sgr data set, which is much larger than previous literature samples with [C/Fe] in Sgr. 

To our sample of carbon measurements in Sgr, we add those of \citet{Chiti19} and \citet{Chiti20Sgr}, who observed metal-poor Sgr stars with the Magellan Echellette (MagE) Spectrograph, measuring [C/Fe] for 4 and 18 targets, respectively. These stars have metallicities in the range $-3.1 \lesssim \FeH \lesssim -1.5$, similarly to the PIGS/AAT range. None of these stars are CEMP according to the standard definition ([C/Fe] $>+0.7$). 
Other Sgr members with measured [C/Fe] that are not included  are the targets analysed in \citet{Hansen18Sgr} and from APOGEE DR17. \citet{Hansen18Sgr} measured [C/Fe] in 12 stars with metallicity $-2.95 \lesssim \FeH \lesssim -1.40$. These targets were observed with UVES high-resolution spectrograph at VLT. However, as shown in \citet{Sestito24Sgr}, the [C/Fe] ratios from \citet{Hansen18Sgr} are systematically lower than the ones from  \citet{Chiti19}, \citet{Chiti20Sgr}, and this work \citep[see Figure~5 in][]{Sestito24Sgr}. APOGEE stars are not included, since the C-measurements are in non-local thermodynamic equilibrium (non-LTE) and in the infra-red, which have offsets compared to LTE measurements in the optical \citep{Jonsson20}.

\subsection{New CEMP stars in Sgr}

The distribution of [Fe/H] vs A(C) for Sgr stars is shown in Figure~\ref{Fig:ac} (blue circles). According to the classical definition of CEMP stars ([C/Fe]~$> +0.7$), only 3-4 stars in the PIGS/AAT Sgr sample are classified as CEMP. One of them (red pentagon) has previously been studied in \citet{Sestito24Sgr}, and was confirmed to be a CEMP-s star based on the over-abundance of s-process elements ([Ba/Fe]~$\sim+1.2$). For the other two CEMP candidates, Ba measurements are not available. We compare the distribution of metallicities and carbon abundances with those for the inner Galaxy (grey circles) and DGs \citep[][orange circles]{Lucchesi24}. Note that the DG sample only includes carbon-normal and spectroscopically confirmed CEMP-no stars. 

\begin{figure}
\includegraphics[width=0.5\textwidth]{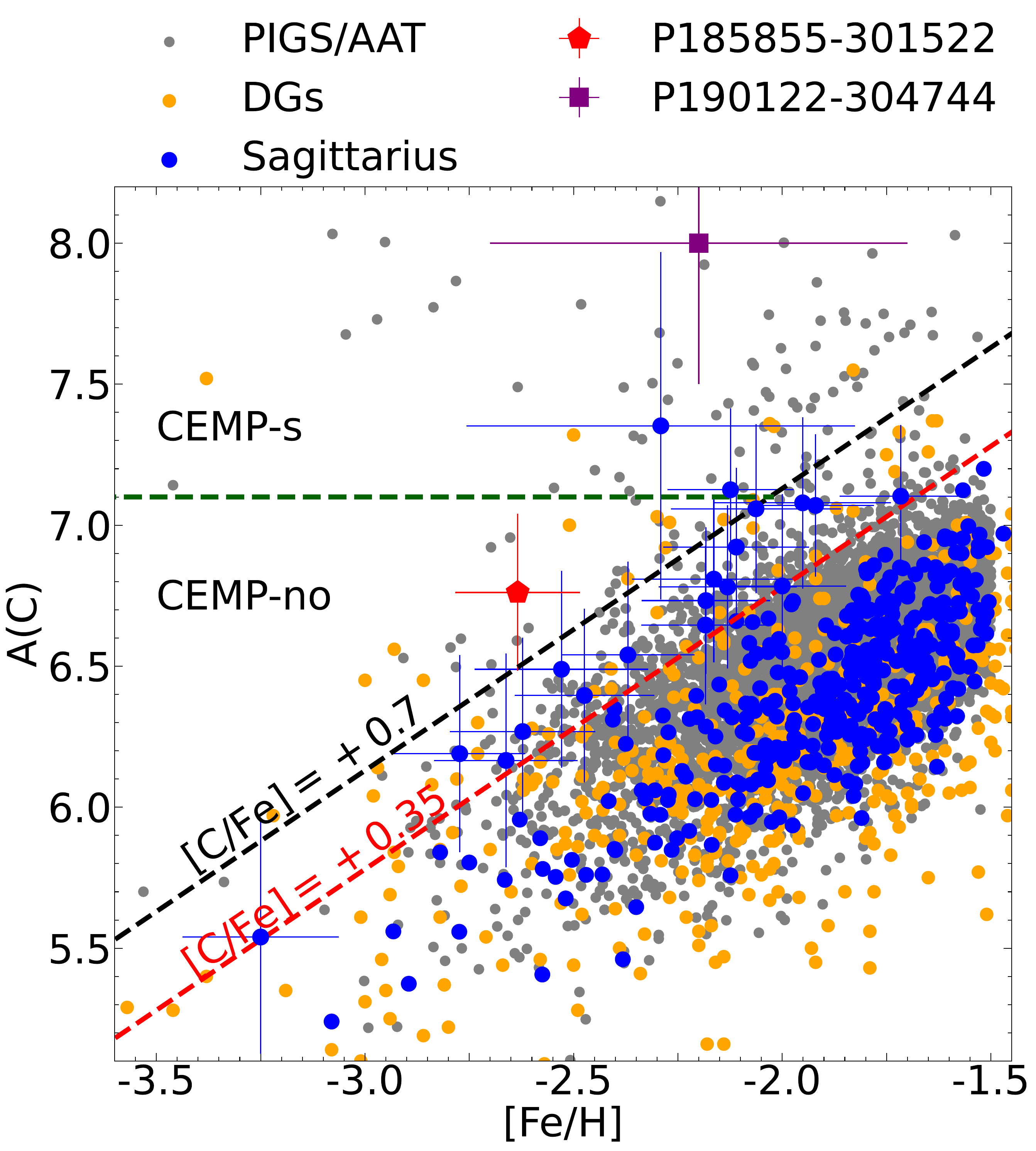}
\caption{Abundance of C, A(C), as as function of  \FeH. The Sagittarius sample (blue circles) includes stars from the final selection made in Section~\ref{sec:sgrselection}, from \citet{Chiti19}, and from \citet{Chiti20Sgr}. The DGs compilation (orange circles) is from \citet{Lucchesi24}. Inner Galaxy stars (grey circles) are selected from \citet{Arentsen24} to have good quality of the AAT spectra and good \texttt{FERRE} measurements as in our sample. Star P185855$-$301522  (red pentagon) is analysed in \citet{Sestito24Sgr} and confirmed to be CEMP-s from high-resolution spectroscopy. Star P190122-304744 (purple square) is one of the two cool CEMP candidates discussed in Section~\ref{sec:weird}. Horizontal green dashed line tentatively separates CEMP-s from CEMP-no as in \citet{Yoon16}. Stars on the left of the  dashed black line have [C/Fe]~$>+0.7$ as defined in \citet{Aoki07}. The dashed red line denotes the tentative new limit for CEMP in Sgr ([C/Fe]~$=+0.35$). Sgr stars with [C/Fe]~$>+0.35$ are displayed with their errorbars  to highlight that they are significantly distant from the bulk of the system's distribution.}
\label{Fig:ac}
\end{figure}

Without measurements of Ba or Sr, it is not possible to classify CEMP stars with certainty, although a rough classification can be made based on [Fe/H] and A(C) alone \citep[e.g.][]{Yoon16}. CEMP-s stars typically have higher A(C) than CEMP-no stars and are more common at higher metallicities, and a tentative separation between the two groups has been placed at A(C)~$= 7.1$ \citep{Yoon16} and [Fe/H]$\gtrsim-3.3$. It is not entirely clean -- there is some known contamination when using such a simple division without detailed chemistry, for example the \citet{Sestito24Sgr} CEMP-s star lies in the CEMP-no region based on [Fe/H] and A(C) alone, and some DG CEMP-no stars lie in the CEMP-s region. Similarly, a contamination of CEMP-no in the CEMP-s region is also found for MW halo stars \citep[\eg][]{Norris19}. However, without better data, we may propose that the two new Sgr CEMP stars are likely of the CEMP-s kind given their metallicity and high carbon abundances.

\subsection{Two cool candidate CEMP stars}\label{sec:weird}

We noticed that there are two stars in the AAT/Sgr sample (not passing our \texttt{FERRE} quality cuts,  based on $\chi^2$) that by eye appear to be very carbon-rich from their spectrum. These stars, Pristine\_185524.38-291422.5 (Gaia DR3 source\_id = 6761678859361894912) and Pristine\_190122.55-304744.3 (6760545743905626496) are highlighted with pink circles in the \textit{Pristine} colour-colour diagram in the top left and right panels of Figure~\ref{Fig:selection}. It is curious that one of them is located above the primary Sgr sequence in the Pristine colour-colour diagram. They are also shown on the CMD with large red symbols in the top panel of Figure~\ref{Fig:weird}. The same star that is an outlier in the colour-colour diagram is located beyond the metal-rich side of the RGB, which is also curious. If the star is truly a Sgr star (and there is no reason to suspect it is not given its radial velocity and proper motions), it cannot be an intrinsic carbon star, because it is not evolved enough. 

Both stars have \texttt{FERRE} $T_\mathrm{eff} \sim 4500$~K, which is at the cool boundary of the \texttt{FERRE} grid, therefore they could be even cooler. Inspection of the spectroscopic fit shows that the \texttt{FERRE} fit is bad in both the blue and the CaT regions: there is a strong discrepancy between the carbon features in the star and those in the FERRE grid, although it is clear that the star is very carbon-rich. This is potentially due to the assumptions on nitrogen in the \texttt{FERRE} grid (see below). 

\begin{figure}
\includegraphics[width=0.45\textwidth]{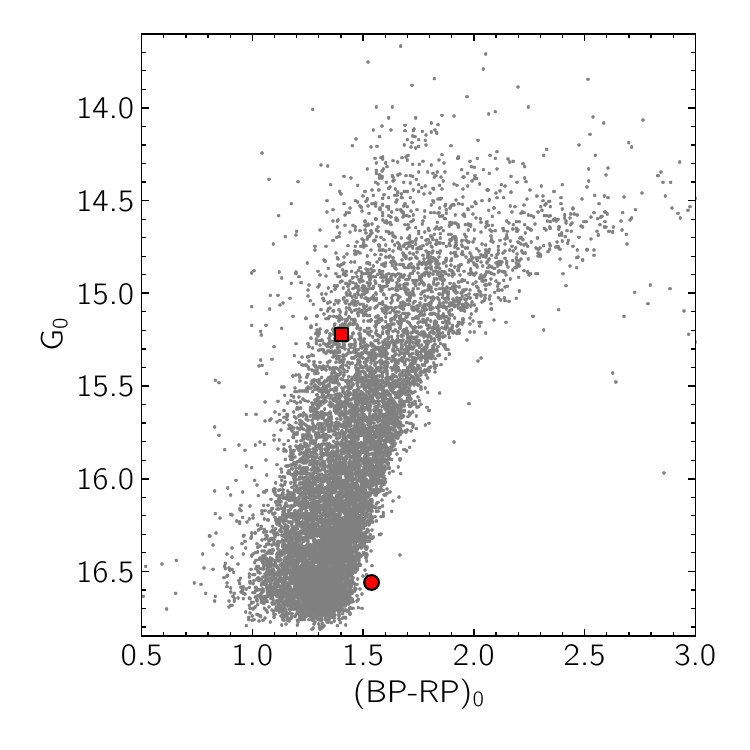}
\includegraphics[width=0.5\textwidth]{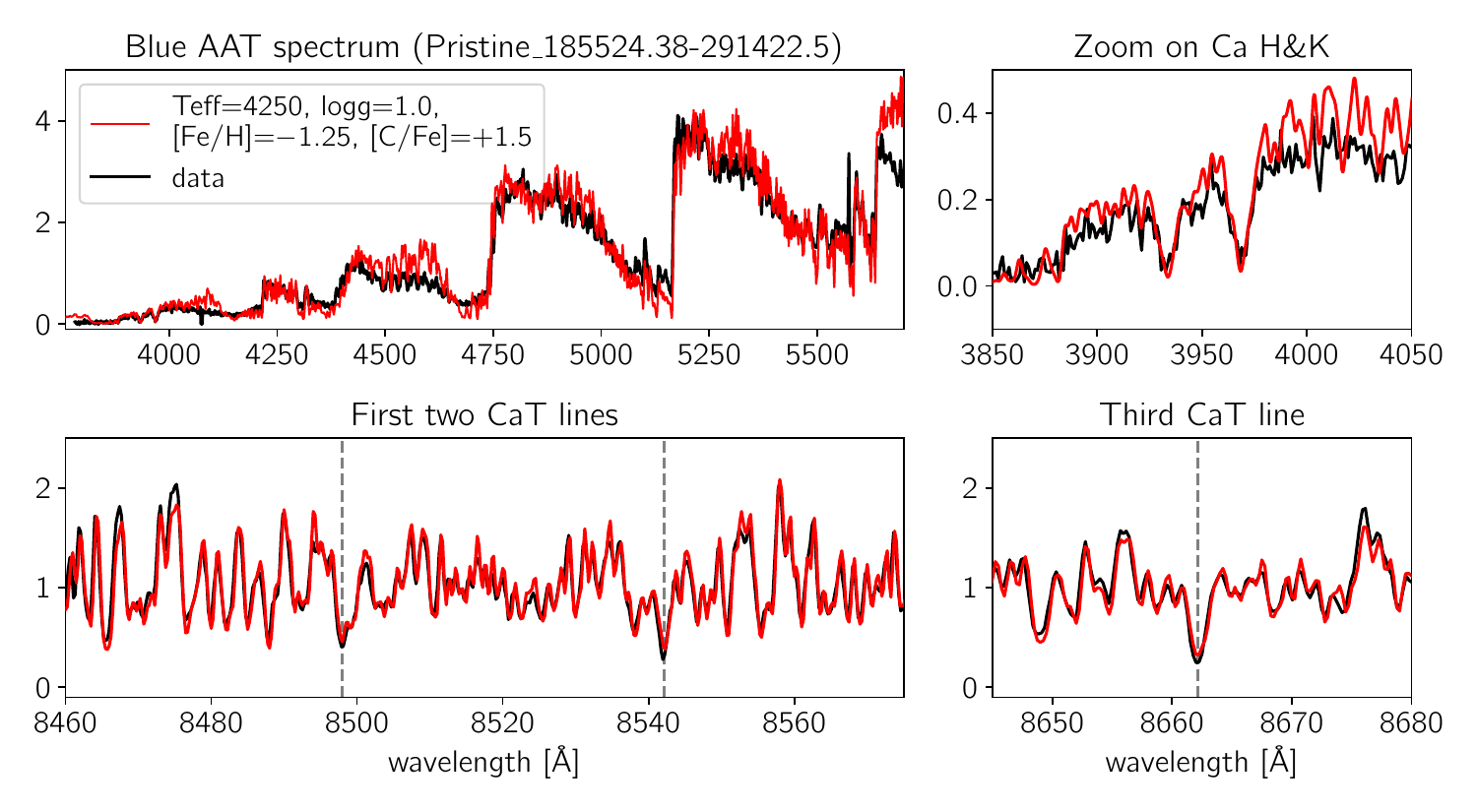}
\includegraphics[width=0.5\textwidth]{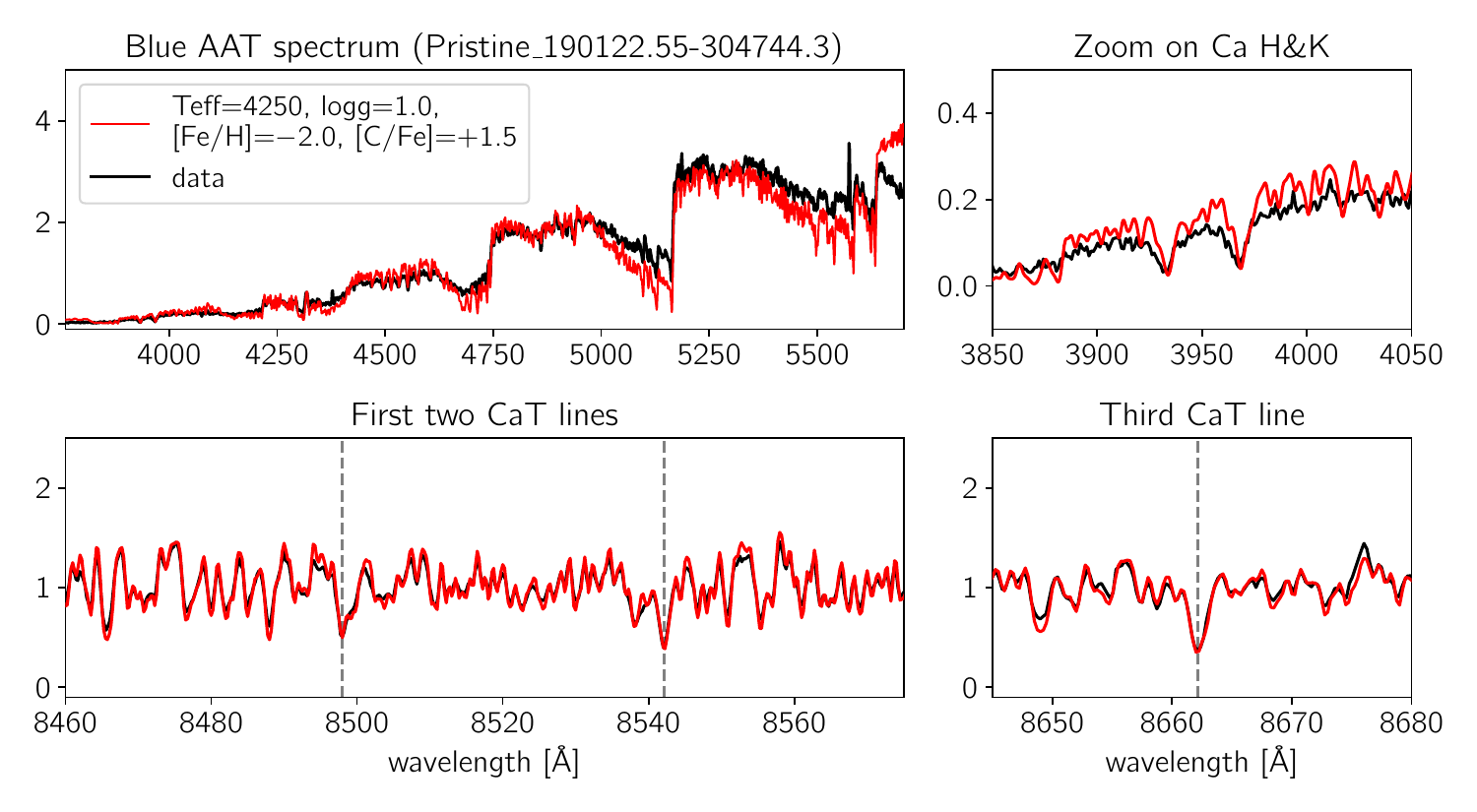}
\caption{Top: colour-magnitude diagram of stars in Sagittarius (same samples as grey dots in Figure~\ref{Fig:selection}), with the CEMP candidates Pristine\_185524.38-291422.5 (circle) and Pristine\_190122.55-304744.3 (square) highlighted with large red symbols. Middle: one of the best matching spectra from the SSPP synthetic grid for Pristine\_185524.38-291422.5 on top of its AAT spectrum. Bottom: same but for Pristine\_190122.55-304744.3.}
\label{Fig:weird}
\end{figure}

To further constrain the stellar parameters for these stars, we employ a different grid of synthetic spectra originally created for use in the Segue Stellar Parameter Pipeline (\texttt{SSPP}, \citealt{Lee08a, Lee08b}, grid from Y.S. Lee, private communication). An important difference between the \texttt{FERRE} and \texttt{SSPP} grids is that the former assumes [N/Fe] = 0, while the latter assumes [C/N] = 0 -- this is potentially particularly important for fitting the CN features in the CaT. We use a cool subset of the grid with the following stellar parameters: T$_\mathrm{eff}$ = [4000, 4250, 4500, 4750]~K, $\log g = 1.0$ (we checked that varying $\log g$ does not make a difference), \FeH{} from $-3.0$ to $-1.0$ in steps of 0.25 dex and [C/Fe] from $0.0$ to $+3.0$ in steps of 0.25 dex. After normalising both the observed and synthetic spectra with a running median of 200 pixels (50 \AA), we search for the best matching spectrum by minimising the residuals. We do this separately for the CaT and the blue and combine the $\chi^2$ values afterwards, giving more weight to the CaT because of its high resolution and because it is less sensitive to the shape of the molecular bands. 

For both of the stars there is no clear best-fit stellar parameter combination, because there are strong degeneracies between T$_\mathrm{eff}$, [Fe/H] and [C/Fe]. For Pristine\_185524.38-291422.5, the outlier in photometry, the main constraint is placed on the absolute carbon abundance: for the $5\%$ best fits, A(C)~$= 8.7 \pm 0.4$ (mean and standard deviation). The mean metallicity is $-1.5 \pm 0.4$ and the temperature is not well-constrained within the limit of our small grid. The other star, Pristine\_190122.55-304744.3, is more metal-poor and slightly less carbon-rich -- the mean A(C) = $8.0 \pm 0.5$ and [Fe/H] = $-2.2 \pm 0.5$ for the $5\%$ best fits, and the temperature is again not well-constrained. For each of these stars, we present one of the best matching synthetic spectra in  Figure~\ref{Fig:weird}, with the observed spectrum in black and the synthetic one in red. We applied a by-eye linear normalisation to the blue arm synthetic spectrum to roughly match the shape of the observed spectrum rather than showing the normalised version, so the match is not perfect.  

We conclude that these stars are likely CH- or CEMP-s stars. The location of the more metal-rich star in the \textit{Pristine} colour-colour diagram and the CMD is likely strongly affected by the very large carbon bands, causing the star to look fainter and redder compared to where a ``normal'' metal-poor star would be. This effect appears to be less strong for the more metal-poor star, although it is on the border of having been included in our selection according to Figure~\ref{Fig:selection}. Such extreme stars have likely been missed in other selections of metal-poor stars as well, in DGs and the Milky Way, possibly leading to an underestimate of the number of binary mass-transfer type stars at intermediate metal-poor metallicities.

\subsection{Fraction of CEMP stars}

In the Galactic halo, the cumulative fraction of CEMP stars for $\FeH < -2.0$ has been found to be of the order of $20-30\%$, rising to $30-40\%$ for $\FeH < -3.0$ \citep{Lee13,Placco14,Arentsen21}. There are various caveats complicating the exact determination of the overall CEMP and separate CEMP-no and CEMP-s fractions in the Galactic halo \citep{Arentsen21}, but the consensus is that there is a significant fraction of these stars at low metallicity. As shown in Figure~\ref{Fig:ac}, only three out of 356 PIGS/AAT Sgr stars is classified as CEMP and none from \citet{Chiti19} and \citet{Chiti20Sgr}, giving a total percentage of $\sim3\%$ for $\FeH < -2.0$ and $\sim5\%$ for $\FeH < -2.5$ -- much lower than that claimed in Galactic halo samples. This could partially be the result of our photometric metal-poor candidate selection being biased against carbon-rich stars, especially those at slightly higher metallicity ($\FeH > -2.5$) and/or higher carbon abundance ([C/Fe]~$>+1.5$) -- the realm of the CEMP-s stars. 

The CEMP fraction in Sgr is also low for $\FeH < -2.5$, and we find that none of the 8 Sgr stars with $\FeH < -2.7$ are CEMP. This is interesting given that in our test of the selection function in Section~\ref{sec:seleffect}, we found that CEMP-no stars in this metallicity range should typically not have been excluded from our selection. This finding is consistent with previous observations suggesting that classical DGs are poor in CEMP-no stars in comparison to the MW and UFDs  \citep[\eg][]{Starkenburg13, Jablonka15, Kirby15, Simon15, Hansen18, Lucchesi24, Skuladottir15, Skuladottir21, Skuladottir24,Chiti24}.

\subsection{Redefining CEMP stars in DGs}

Given that the average carbon abundance is $\sim0.3$~dex lower in Sgr compared to the Milky Way (Figure~\ref{Fig:cfe_comp}), is it fair to use the same definition of carbon-enhancement as in the Milky Way? This seems to be a generic question for classical DGs, as most of them have lower average [C/Fe] than the Milky Way, as discussed in the previous section, and they would therefore need a larger carbon ``boost'' to be classified as CEMP. The LMC also has lower carbon abundances compared to the MW halo (although similar to the inner Galaxy), with a dearth of CEMP stars \citep{Jonsson20,Chiti24,Oh24}. The first definition of CEMP stars was [C/Fe]~$>+1.0$ \citep{Beers05}, which was refined empirically by \citet{Aoki07} to [C/Fe]~$>+0.7$ based on a sample of observations of MW stars, using the gap between carbon-normal stars and outliers with high carbon abundances. This definition is therefore a relative one, specifically for the Milky Way ``field'' population, raising the question whether it should it be redefined for dwarf galaxies. 

Inspecting Figure~\ref{Fig:ac}, there are a significant number of Sgr stars that appear to be outliers in A(C) from the main Sgr trend, although they do not make it to above the classical CEMP definition of [C/Fe]~$>+0.7$. For $\FeH\lesssim-2.5$ in the PIGS/AAT inner Galaxy sample, the average [C/Fe]~$\approx +0.3$ with a dispersion of 0.2~dex (conservative estimate) -- meaning that the [C/Fe]~$=+0.7$ CEMP definition selects stars that are $\sim 2 \sigma$ outliers, roughly 0.4~dex higher than the mean trend. The average [C/Fe] in Sgr in the lowest metallicity bins ($\FeH \lesssim -2.5$) is $\sim-0.05$, therefore, adopting a similar conservative dispersion, stars with [C/Fe] $> -0.05 + 0.4 > +0.35$ could be considered outliers in Sgr, and therefore CEMP. This working definition of CEMP stars in Sgr is shown in Figure~\ref{Fig:ac} with a dashed red line. Using this new definition, $\sim20$  Sgr members would be classified as CEMP stars (vs $3-4$ from the classical definition). This would lead to a carbon-enhanced percentage of $\sim15\%$ for $\FeH < -2.0$, which is much less in tension with the results in the MW \citep[$20-30\%$,][]{Lee13,Placco14,Arentsen21}. The percentage would be $\sim12\%$ for $-2.5<\FeH < -2.0$ and $\sim30\%$ for $\FeH < -2.5$ (or $\sim35\%$ if only Sgr/AAT data are considered), compatible with the frequency of CEMP stars in the MW.

Similarly, for Dra, UMi, and Scl (selecting stars between $-2.4<\FeH<-1.9$), the new [C/Fe] threshold for a member star to be a CEMP would be $\sim+0.3,+0.3,+0.1$, respectively. This new limit would suggest that the percentage of CEMP in Dra, UMi, and Scl would be $\sim16\%, 27\%, 19\%$, respectively. However, the latter values refer only to the CEMP-no population, given that the compilation from \citet{Lucchesi24} does not contain CEMP-s stars.

We want to highlight that our new definition of CEMP is strictly empirical and based on the position of outliers in the [C/Fe] or A(C) distribution -- they could be enhanced in carbon for a number of reasons. A more physically driven definition should take into account the IMF and the energy ranges of the SNe~II exploded in a given system, the contribution of SN~Ia and AGBs, the binary fraction, and the efficiency of the system's ISM in recycling the ejected yields. Additionally, investigations of the chemical properties of CEMP candidates based on our relative CEMP definition  will be necessary to test whether they show differences in their abundance patterns compared to stars in the bulk of the carbon-metallicity distribution, and whether they are truly a different population of stars.

\section{Summary}\label{sec:conclusions}
The chemo-dynamical properties of the low-metallicity regime of the Sagittarius dwarf galaxy are explored using the low/medium-resolution AAT spectra observed by the \textit{Pristine} Inner Galaxy Survey (PIGS). The PIGS dataset contains measurements of RVs, stellar parameters, \FeH{}, and [C/Fe] for stars towards the inner Galaxy and Sgr. We summarise below the main conclusions of this work: 

\begin{enumerate}[I)]
    \item We provide a clean list of low-metallicity ($\FeH\leq-1.5$) members stars selected according to their RV from AAT and proper motion and on-sky position from Gaia, as in \citet{Vitali22}, and updated to DR3 (Figure~\ref{Fig:onsky}). A table updated to Gaia DR3 will be available as online material. 
    \item The metal-poor ($\FeH\leq-1.5$) population (PIGS/AAT) of Sgr has a larger velocity dispersion and systemic RV than the metal-rich ($\FeH\geq-0.6$, APOGEE) as shown in Figures~\ref{Fig:hist}~and~\ref{Fig:disp}. Additionally,  the velocity dispersion and the systemic RV increase in the outer regions for both  populations. This effect might be caused by the contribution of various mechanisms, such as the complex structure in Sgr (MR/disc $+$ MP/halo), the outside-in star formation, and the extreme Galactic tidal perturbations acting in the system.
    \item The average [C/Fe] of Sgr is similar to the range displayed by the other classical DGs (Figure~\ref{Fig:cfe_comp}). However, the level of [C/Fe] is higher in Sgr than in Car and Scl. This can be explained by differences in the IMF and in the energy distribution of the SNe~II among these systems, with a predominance of more energetic events in Car and Scl.
    \item The average [C/Fe] of Sgr, and of the other classical DGs, is lower than in the MW at fixed $\FeH$, either when compared to inner Galactic  or halo-like stars (Figure~\ref{Fig:cfe_comp}). The ISM of classical DGs might have been able to retain the ejecta of energetic events, such as hypernovae, while this would not have been the case for the  building blocks of the Galaxy, where stochasticity might have played an important role. In this scenario, classical DGs should display the imprint of Population~III~and~II high energy  SNe~II, which would act to lower the average [C/Fe]. Instead, less energetic events, faint- and core-collapse SNe~II from a more pristine population should be imprinted in the stars of the MW building blocks, hence the higher [C/Fe]. On the other hand, some studies \citep[\eg][]{Deason16} suggest that the majority of the MW building blocks should be similar in size to present DGs. However, their chemical evolution still remain an open question. Our results indicate a different supernovae imprint between Sgr (and classical DGs) vs the MW building blocks.
    \item SNe~Ia can also lower the average [C/Fe]. This kind of event would be already present at $\FeH\sim-2.0$ in classical DGs and absent in the MW stars at the same metallicities. Indications of the SNe~Ia contributions in Sgr starting at $-2.0<\FeH<-1.5$ are the lower  median [C/Fe] at these metallicities vs the higher  [C/Fe] at lower metallicities (see Figure~\ref{Fig:cfe_comp}) and also the lower [C/Fe] in the inner regions (see Figure~\ref{Fig:cfe_ell}), inhabited by a slightly more metal-rich population. The presence of SNe~Ia at the aforementioned metallicities would also be confirmed by the trend of [Co/Fe]  found by \citet{Sestito24Sgr}.
    \item We find a positive [C/Fe] gradient of $\nabla \rm{[C/Fe]} \sim 0.23\ \rm{dex \ r_h^{-1}}$ or $ \sim 8.8\times 10^{-2}\ \rm{dex \ kpc^{-1}}$ or $\sim 6.8\times 10^{-4}\ \rm{dex \ arcmin^{-1}}$ for stars with $-2.0<\FeH<-1.5$ (Figure~\ref{Fig:cfe_ell}), which we interpret as the effect of contributions by SNe~Ia. 
    \item We  identify $4$ new CEMP stars in Sgr. Figure~\ref{Fig:ac} suggests that the empirical distinction between CEMP-s and CEMP-no solely based on A(C) does not work well for Sgr and the classical DGs. We therefore cannot reach definitive conclusions on the nature of the new CEMP stars, however, we may propose that they are likely of the CEMP-s kind given their \FeH{} and high A(C). 
    \item The AAT spectra of two carbon-rich candidates, Pristine\_185524.38-291422.5 and Pristine\_190122.55-304744.3, are re-analysed with the \texttt{SSPP} grid of synthetic spectra (Figure~\ref{Fig:weird}) because they had high $\chi^2$ in the \texttt{FERRE} fit and were at the edge of the \texttt{FERRE} grid. They have [Fe/H]~$\sim -1.5$ and $-2.2$ with very high carbon abundances (A(C)~$\sim 8.8$ and $8.0$, respectively), making them  CH- or CEMP-s candidates. The C-bands of the former star strongly affect its colour, magnitude and its position in the \textit{Pristine} colour-colour diagram (Figures~\ref{Fig:selection} and \ref{Fig:weird}). Similar stars could have been missed in other metal-poor (DG) selections as well.
    \item The photometric selection effects in the various PIGS fields that include Sgr targets are discussed, showing there is a bias against CEMP stars in the sample (Figure~\ref{Fig:selection}),  specifically those of the CEMP-s (binary interaction) type. CEMP-no stars (connected to early chemical evolution), however, are less likely to have been excluded from the selection and their frequency in our sample should be largely unbiased.
    \item Following the classical definition of CEMP stars ([C/Fe]~$>+0.7$), the fraction of CEMP stars in our sample is very low: $\sim3\%$ for $\FeH < -2.0$ and $\sim6\%$ for $\FeH < -2.5$. However, the low mean abundance of [C/Fe] in Sgr (and other classical DGs) as well as the clear presence of outliers of the distribution at ``intermediate'' carbon abundances, lead us to propose a new definition for CEMP stars. Rather than a fixed threshold, the limit should depend on the average [C/Fe] of a given system. For Sgr, stars with [C/Fe]~$\gtrsim+0.35$ can be considered CEMP in this case, as they are outliers from the bulk of the system's distribution (see Figure~\ref{Fig:ac}). The new frequency of CEMP in Sgr according to this definition would be $\sim12\%$ for $-2.5 < \FeH < -2.0$ and $\sim30-35\%$ for $\FeH < -2.5$, much more in agreement with frequencies in the MW.
    
\end{enumerate}

This work, which complements the high-resolution investigation by \citet{Sestito24Sgr}, provides a novel glimpse into the early chemical evolution of Sgr by exploring its level of carbon. These works will be beneficial for upcoming spectroscopic surveys, for example 4DWARFS \citep{Skuladottir23}, which will observe a larger number of stars in the Sgr core and in its streams.

\begin{acknowledgements}
We acknowledge and respect the l\textschwa\textvbaraccent {k}$^{\rm w}$\textschwa\ng{}\textschwa n peoples on whose traditional territory the University of Victoria stands and the Songhees, Esquimalt and WS\'ANE\'C  peoples whose historical relationships with the land continue to this day.
\\

We thank the Australian Astronomical Observatory, which have made the PIGS spectroscopic follow-up observations used in this work possible. We acknowledge the traditional owners of the land on which the AAT stands, the Gamilaraay people, and pay our respects to elders past and present. 
\\

We want to thank the anonymous referee for their helpful and insightful comments.
\\

We thank Vini Placco for calculating the carbon evolutionary corrections. 
We thank Young Sun Lee for providing the SSPP synthetic spectra. 
\\

FS and KAV thank the National Sciences and Engineering Research Council of Canada for funding through the Discovery Grants and CREATE programs. AAA acknowledges support from the Herchel Smith Fellowship at the University of Cambridge and a Fitzwilliam College research fellowship supported by the Isaac Newton Trust. SV thanks ANID (Beca Doctorado Nacional, folio 21220489) and Universidad Diego Portales for the financial support provided. SV acknowledges the Millennium Nucleus ERIS (ERIS NCN2021017) and FONDECYT (Regular number 1231057) for the funding. NFM gratefully acknowledges support from the French National Research Agency (ANR) funded project ``Pristine'' (ANR-18-CE31-0017) along with funding from the European Research Council (ERC) under the European Unions Horizon 2020 research and innovation programme (grant agreement No. 834148). ES acknowledges funding through VIDI grant ``Pushing Galactic Archaeology to its limits'' (with project number VI.Vidi.193.093) which is funded by the Dutch Research Council (NWO). This research has been partially funded from a Spinoza award by NWO (SPI 78-411). 
\\

The spectroscopic follow-up used in this work was based on selection from observations obtained with MegaPrime/MegaCam, a joint project of CFHT and CEA/DAPNIA, at the Canada–France–Hawaii Telescope (CFHT) which is operated by the National Research Council (NRC) of Canada, the Institut National des Science
de l'Univers of the Centre National de la Recherche Scientifique (CNRS) of France, and the University of Hawaii.
\\

This work has made use of data from the European Space Agency (ESA) mission {\it Gaia} (\url{https://www.cosmos.esa.int/gaia}), processed by the {\it Gaia} Data Processing and Analysis Consortium (DPAC, \url{https://www.cosmos.esa.int/web/gaia/dpac/consortium}). Funding for the DPAC has been provided by national institutions, in particular the institutions participating in the {\it Gaia} Multilateral Agreement.
\\

This research has made use of the SIMBAD database, operated
at CDS, Strasbourg, France \citep{Wenger00}. This work made
extensive use of TOPCAT \citep{Taylor05}.
\\
\\
\textbf{Author contribution statement.} FS led the analysis and the various discussions in this work, contributed to write most of this draft, and created most of the Figures. AAA led the PIGS/AAT target selection and observations, co-led the PIGS/AAT spectroscopic analysis with David Aguado (not a co-author here), analysed the cool candidate CEMP stars discussed in Section~\ref{sec:weird}, created the respective section with figures, and was closely involved in shaping the manuscript and contributed to the scientific discussion. SV contributed to the discussion and revision of the paper. MM identified one of the cool candidate CEMP stars using photometry and contributed to the discussion. RL provided the [C/Fe] dataset of the various dwarf galaxies. KAV, NFM, JFN, and ES provided insightful scientific and editorial comments on the manuscript.

\end{acknowledgements}

\bibliographystyle{aa}
\bibliography{Sgr_C_map}

\end{document}